\documentclass[preprint,12pt]{elsarticle}

\usepackage[T1]{fontenc}
\usepackage[utf8]{inputenc}
\usepackage{amsmath,amssymb}
\usepackage{graphicx}
\usepackage{booktabs}
\usepackage{lmodern}
\usepackage[hidelinks,hypertexnames=false]{hyperref}
\usepackage{float}
\usepackage{tabularx}
\usepackage{longtable}
\usepackage{tikz}
\usetikzlibrary{arrows.meta, positioning, fit}

\journal{Computer Physics Communications}

\begin{document}

\begin{frontmatter}

\title{AloHEP Software: A luminosity optimizer for High Energy Physics colliders}

\author{B. Da\u{g}l\i}
\ead{b.dagli@etu.edu.tr}
\author{Y. Karaca\corref{cor1}}
\ead{ykaraca@etu.edu.tr}
\cortext[cor1]{Corresponding author.}
\address{TOBB University of Economics and Technology, Ankara, T\"urkiye}
\begin{abstract}
The luminosity quoted for proposed particle colliders is usually derived from
the analytic geometric formula, which assumes head-on collisions of identical, rigid
Gaussian bunches. For the new generation of asymmetric machines---in
particular lepton--hadron colliders such as the EIC, LHeC and FCC-eh---this assumption
can misestimate the luminosity, because it ignores a few coupled effects
that act in the interaction region such as hourglass effect, crossing angle, and mutual
electromagnetic (beam--beam) deformation of the colliding bunches. We present \textbf{AloHEP}
(A luminosity optimizer for High Energy Physics), an open-source, cross-platform Java
application that simulates the interaction region of both symmetric and asymmetric colliders.
AloHEP represents each bunch by an ensemble of macroparticles deposited on the grid with a
cloud-in-cell scheme, and slices the crossing longitudinally. On each transverse slice it
solves the two-dimensional electrostatic problem by a direct Green's-function convolution,
then obtains the beam--beam forces by finite differencing and continuous bicubic interpolation. The hourglass effect, finite crossing angle and beam--beam pinch/disruption can be enabled independently, and the two colliding beams may consist of entirely different species (leptons, protons, heavy ions). A JSON-backed parameter database and an interactive graphical interface make AloHEP a flexible tool for the rapid prototyping of realistic luminosity estimates. We verify that the code reproduces the analytic geometric luminosity in the appropriate limit
and present scaling, grid-convergence and asymmetric-collider benchmarks.
\end{abstract}


\end{frontmatter}

%================================================================
\section*{Program summary}
\noindent
\begin{small}
\begin{tabular}{@{}p{0.32\linewidth}p{0.62\linewidth}@{}}
\textit{Program title:} & AloHEP \\
\textit{CPC Library link to program files:} & (to be added by Technical Editor) \\
\textit{Developer's repository link:} & \url{https://github.com/tobbetuyef/AloHEP} \\
\textit{Licensing provisions:} & MIT License (the bundled Gson library is distributed under the Apache License 2.0) \\
\textit{Programming language:} & Java (Java~7-compatible syntax; builds and runs on OpenJDK~8--21, benchmarks run with OpenJDK~17; external library Gson~2.9.1) \\
\textit{Supplementary material:} & Benchmark driver scripts and the raw CSV results of Section~\ref{sec:benchmarks} are provided in the repository. \\
\end{tabular}
\end{small}

\paragraph{Nature of problem}
The analytic geometric luminosity does not reflect realistic operating values, especially
in asymmetric lepton--hadron designs. The hourglass effect, crossing angles and beam--beam
interactions all change the beam cross-section during the collision, so realistic estimates
require high-resolution numerical simulations that can handle different particle species
(e.g.\ electrons, protons, heavy ions) and collider types (ring-ring, linac-linac, linac-ring). Together with the luminosity, AloHEP reports for each beam the beam--beam tune shifts
$\xi_{x,y}$ (stored beams) or the disruption parameters $D_{x,y}$ (single-pass beams), so
that a design can be checked against the beam--beam limits. Let us emphasize that, a luminosity quoted without the beam--beam tune shifts of the stored beams (or the
disruption of single-pass linac beams) can be unrealistically high, because these grow with
the population and density of the opposing bunch and must stay below their stability limits.

\paragraph{Solution method}
AloHEP uses a Particle-in-Cell (PIC) scheme with cloud-in-cell charge assignment, reducing
each three-dimensional bunch crossing to a sequence of overlapping two-dimensional transverse
slices. The potential of each slice is obtained by direct convolution of the gridded charge
density with the two-dimensional logarithmic Green's function, and the forces by finite
differencing of the potential (evaluated at the cell faces, midway between potential grid
points). The macroparticle forces are then reconstructed with a $C^{1}$ bicubic
(Catmull--Rom~\cite{catmull}) interpolation, which avoids the gradient discontinuities at the
cell boundaries of bilinear ($C^{0}$) interpolation. Both colliding beams may differ in species,
energy, emittance and population. A preliminary first-order beamstrahlung model for light particles (the average quantum-corrected energy loss~\cite{yokoyachen}) is also included; it is not yet validated and is not used in the benchmarks of this paper (Section~\ref{sec:beamstrahlung}).

\paragraph{Additional comments including Restrictions and Unusual features}
The hourglass effect, crossing angle and beam--beam effects can each be toggled independently,
and an interactive interface with real-time macroparticle animation lets the bunch
deformation be inspected visually. All
accelerator parameters are stored in a human-readable JSON database, so new machine designs
or different pairings of the built-in species can be set up without recompilation. The runtime is dominated by the direct Poisson solve,
$\mathcal{O}(N_x^{2}N_y^{2})$ per slice repeated $\mathcal{O}(N_z^{2})$ times per
crossing, i.e.\ $\mathcal{O}(N_g^{6})$ for a cubic grid
(Section~\ref{sec:scaling}), and is essentially independent of the macroparticle count up
to $\sim\!10^{5}$. On one workstation core, a full crossing of
$5\times10^{4}$ macroparticles on a $30^{3}$ grid completes in $\sim\!12$~s and a $100^{3}$
convergence run in $\sim\!3.2$~h within a 4~GB Java heap; no HPC environment is required. The simulation is restricted to ultra-relativistic beams
($v\approx c$), small crossing angles, a single bunch-pair crossing (no multi-bunch effects)
and Gaussian initial distributions; macroparticles are advanced by first-order (Euler)
integration, beamstrahlung is included only at the level of the average quantum-corrected
energy loss, and the strongly disrupted regime ($D_y\gg1$) is not validated
(Section~\ref{sec:pinch}).

%================================================================
\section{Introduction}\label{sec:intro}

The scientific reach of a particle collider is set by its centre-of-mass energy and its
luminosity---the factor relating the rate of a given process to its cross-section. A reliable
luminosity estimate is therefore central to every collider proposal. In the design phase this
estimate is almost always taken from the analytic \emph{geometric} luminosity, which assumes
that two identical Gaussian bunches collide head-on and keep their shape throughout the
crossing. Convenient as it is, this assumption ignores three coupled effects that act in the
interaction region: the \emph{hourglass effect}, in which the strong final focusing makes the
transverse beam size vary along the collision axis; the finite \emph{crossing angle},
introduced to suppress parasitic encounters; and the \emph{beam--beam interaction}, the mutual
electromagnetic focusing or defocusing (pinch/disruption) of the colliding bunches, which for
light particles is also accompanied by beamstrahlung. Because these effects are coupled, their
combined impact on the luminosity cannot generally be written in closed form and must be
evaluated numerically.

This matters most for \emph{asymmetric} colliders. A new generation of electron--hadron
colliders aims to deliver far higher luminosity, and a wider range of energies and beam
species, than HERA~\cite{hera}---so far the only electron--hadron collider ever operated. The principal
designs under consideration are the Electron-Ion Collider (EIC), under construction at
BNL~\cite{eic}, the LHC-based Large Hadron electron Collider (LHeC)~\cite{lhec} and muon-hadron colliders~\cite{kaya,akturk2026}  at CERN and the Future
Circular Collider eh option (FCC-eh)~\cite{fcceh,fcchh}. In these machines the two beams differ
in species, energy, emittance and bunch population, so the colliding bunches are usually of
unequal transverse size and the symmetric-collision assumptions behind the geometric formula are not applicable. The need to quantify the hourglass reduction for \emph{unequal} beams was
in fact recognised early in the development of asymmetric machines~\cite{furman}.

Detailed beam--beam simulation has a long and successful history in the linear-collider
community. The two principal Particle-in-Cell tools used for this purpose, CAIN~\cite{cain} and GUINEA-PIG~\cite{guineapig}, describe the colliding bunches through an
ensemble of macroparticles on a three-dimensional grid; at each step the bunches are sliced
longitudinally and the Poisson equation is solved per slice to update the electrostatic
potential, which then governs the particle dynamics. These codes are mature and highly
capable, but they were developed primarily around $e^{+}e^{-}$ linear-collider scenarios,
are written in compiled languages tightly coupled to specific build environments, and are
driven through text-based steering cards---a workflow optimised for production studies
rather than for rapidly exploring the parameter space of a \emph{new} and possibly highly
asymmetric machine.

The present work addresses this gap. We introduce \textbf{AloHEP} (A Luminosity Optimizer
for High Energy Physics), an open-source application written entirely in Java for
cross-platform portability, that simulates the interaction region of arbitrary---and in
particular asymmetric---colliders. An early version of the program was announced in a
conference contribution~\cite{AloHEPipac}, and it has since been used in a number of
lepton--hadron collider design
studies~\cite{acar2017,canbay2017,ketenoglu2022,akay2024,akturk2024,akturk2026}; the
present paper gives the first complete description of the numerical methods, the software
architecture and the verification of the current version. AloHEP uses the same well-established slice-based PIC strategy as CAIN and GUINEA-PIG, but couples it to (i) a dynamic, JSON-backed database that lets any pair of particle species and accelerator parameter sets be
combined and modified without recompilation, and (ii) an interactive graphical interface
with real-time visualisation of the colliding bunches. AloHEP is positioned not as a high-performance production competitor to existing tools, but as a flexible prototyping environment in which the realistic luminosity of a proposed asymmetric collider---together with the associated beam-size, disruption and tune-shift parameters---can be assessed in minutes on a workstation machine. The remainder of this paper presents the physical model of the interaction region (Section~\ref{sec:analysis}), the numerical methods (Section~\ref{sec:numeric}), the software architecture (Section~\ref{sec:software}), and benchmark and verification results (Section~\ref{sec:benchmarks}).

%================================================================
\section{Analysis of the collision region}\label{sec:analysis}

\begin{figure}[htbp]
\centering
\includegraphics[width=0.6\linewidth]{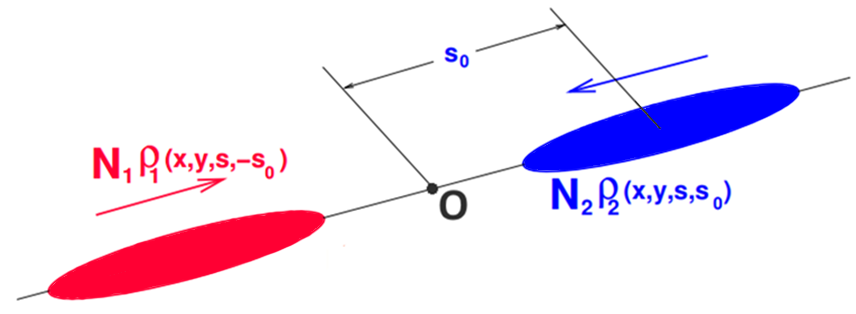}
\caption{The illustration represents the pre-collision positions of opposing bunches. Here
$N$ signifies the number of particles in these bunches, $\rho$ denotes the density
functions, and $O$ is the collision point. The directional axes of bunch movement are
denoted $s$; in a head-on collision the $s$-axis aligns with the $z$-axis. The variable
$s_0$ is the distance of the bunches from the collision centre, corresponding to $c\cdot t$
for particles moving at high energy.}
\label{fig:bunches}
\end{figure}

Colliding two beams is more involved than firing one beam at a fixed target: each beam acts
as both projectile and target, so the density distribution of both bunches matters. A
schematic view of two bunches just before collision is given in Fig.~\ref{fig:bunches}.

Considering a single bunch and the particles it contains, we can define the density
function of the bunch with the following integral:
\begin{equation}
\int \rho\left( x,y,s,s_{0} \right)\,dx\,dy\,ds = 1 .
\end{equation}

By defining the bunches in this manner, it becomes inclusive of situations in which two
bunches exhibit distinct particle distributions and contain different numbers of particles.
Since these two bunches move through each other, the overlap integral depends on the
position of the bunches in the direction of motion and the intersecting time. The
relationship between the overlap integral and luminosity $\mathcal{L}$ can be
expressed as~\cite{herrmuratori}
\begin{equation}
\mathcal{L} \propto K \cdot \int\!\!\int\!\!\int\!\!\int_{-\infty}^{+\infty}
\rho_{1}\!\left( x,y,s,-s_{0} \right)\rho_{2}\!\left( x,y,s,+s_{0} \right)\,dx\,dy\,ds\,ds_{0}.
\end{equation}

The kinematic factor for the collision of beams moving at velocities approaching the speed
of light is
\begin{equation}
K = \sqrt{\left( \vec{v}_{1} - \vec{v}_{2} \right)^{2}
- \left( \vec{v}_{1} \times \vec{v}_{2} \right)^{2}/c^{2}} .
\end{equation}

Under the assumption that the high-energy particle beams travel at the speed of light and
the collision occurs head-on ($v_{1} = -v_{2}$), the kinematic factor reduces to $2c$, i.e.\ $K/c = 2$. Moreover, assuming that the particle density functions are uncorrelated across distinct axes, we may decompose the density distributions, and the luminosity becomes
\begin{equation}\label{eq:lumfull}
\begin{split}
\mathcal{L} = 2N_{1}N_{2}f_{c}\int\!\!\int\!\!\int\!\!\int
&\rho_{1x}(x)\rho_{1y}(y)\rho_{1s}\!\left( s - s_{0} \right)\\
&\times \rho_{2x}(x)\rho_{2y}(y)\rho_{2s}\!\left( s + s_{0} \right)\,dx\,dy\,ds\,ds_{0}.
\end{split}
\end{equation}

Here $N_{1}$ and $N_{2}$ represent the number of particles within the bunches, while
$f_{c}$ denotes the bunch collision rate per second. Given that the particles are
predominantly clustered near the centre, we assume a normal (Gaussian) distribution in all
dimensions:
\begin{equation}
f(x) = \frac{1}{\sigma\sqrt{2\pi}}\, e^{-\frac{1}{2}\left( \frac{x - \mu}{\sigma} \right)^{2}} .
\end{equation}

\begin{figure}[htbp]
\centering
\includegraphics[width=0.55\linewidth]{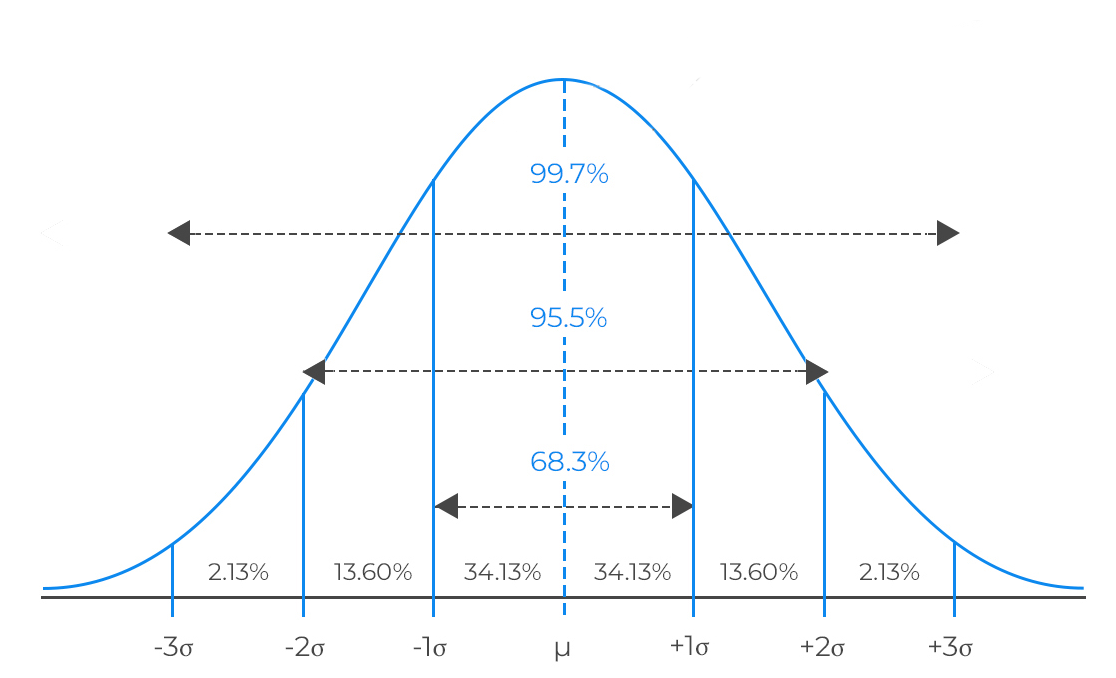}
\caption{The normal distribution function in one dimension. The $1\sigma$ range contains
$68.3\%$, the $2\sigma$ range $95.5\%$, and the $3\sigma$ range $99.7\%$ of the particles in
each dimension.}
\label{fig:normal}
\end{figure}

The mean $\mu$ determines the central position, while the standard deviation $\sigma$
corresponds to its spread. The normal distribution function is shown in
Fig.~\ref{fig:normal}.

The normal distribution functions of particles for each dimension are
\begin{equation}
\begin{aligned}
\rho_{x}(x) &= \frac{1}{\sigma_{x}\sqrt{2\pi}}\exp\!\left( -\frac{x^{2}}{2\sigma_{x}^{2}} \right), \\
\rho_{y}(y) &= \frac{1}{\sigma_{y}\sqrt{2\pi}}\exp\!\left( -\frac{y^{2}}{2\sigma_{y}^{2}} \right), \\
\rho_{s}\!\left( s \pm s_{0} \right) &= \frac{1}{\sigma_{s}\sqrt{2\pi}}
\exp\!\left( -\frac{\left( s \pm s_{0} \right)^{2}}{2\sigma_{s}^{2}} \right).
\end{aligned}
\end{equation}

\begin{figure}[htbp]
\centering
\includegraphics[width=0.45\linewidth]{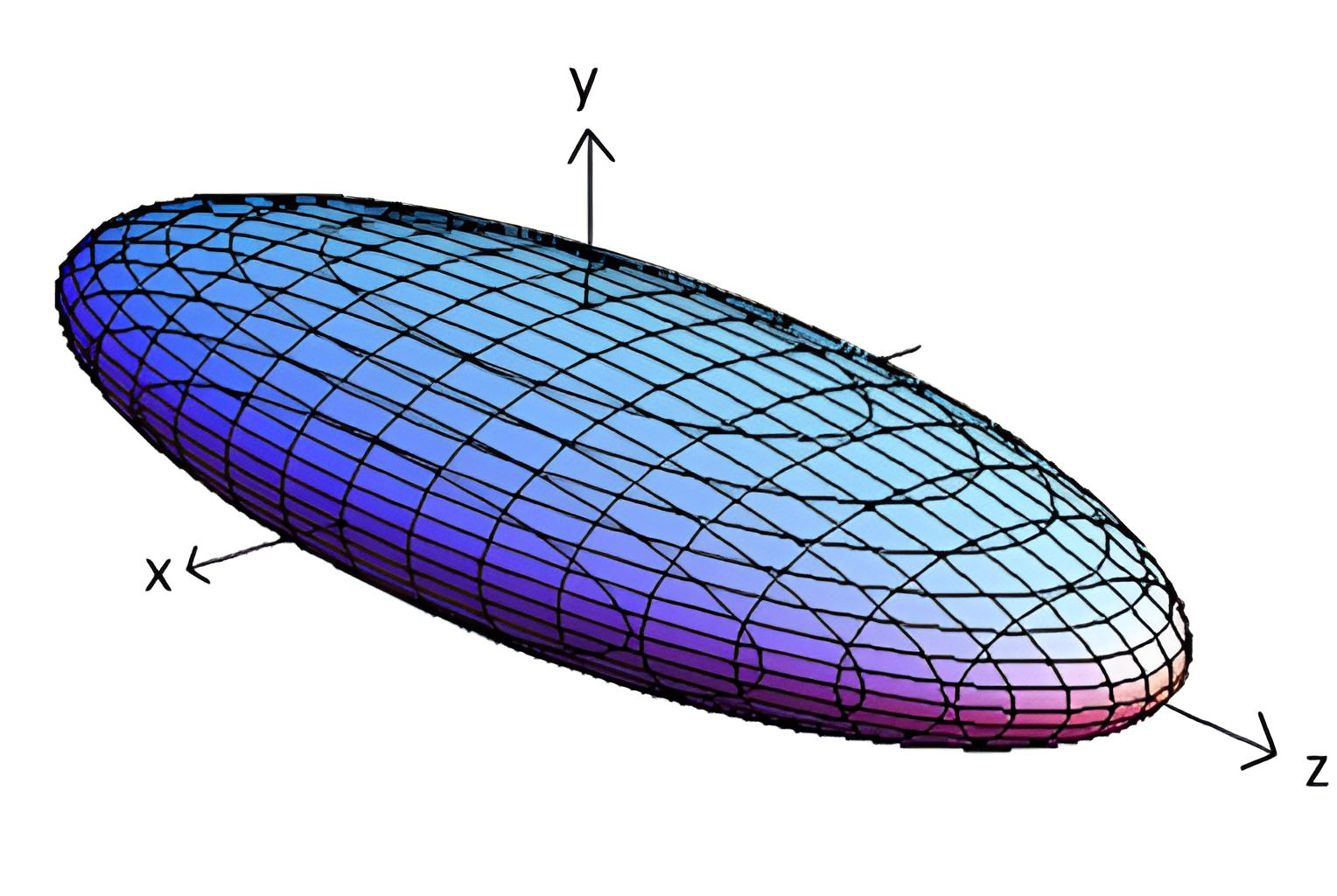}
\caption{Schematic illustration of the particle bunch. The transverse sizes at the
interaction region may be identical or different, depending on the design; the longitudinal
length is significantly greater than the transverse sizes. Beams with equal transverse sizes
are round beams, those with differing sizes are flat beams.}
\label{fig:ellipsoid}
\end{figure}

The centre of the bunch is at zero in the transverse axes and at $\pm s_{0}$ in the
longitudinal axis, and the bunch sizes are $\sigma_{x}$, $\sigma_{y}$ and $\sigma_{s}$.
Consequently the bunch resembles an ellipsoid (Fig.~\ref{fig:ellipsoid}).

Presuming head-on collisions between bunches of identical transverse sizes and no change in
the bunch sizes throughout the collision, integration of Eq.~\eqref{eq:lumfull} yields the
geometric luminosity:
\begin{equation}\label{eq:lumgeo0}
\mathcal{L}_{0} = \frac{N_{1}N_{2}f_{c}}{4\pi\sigma_{x}\sigma_{y}} .
\end{equation}

However, factors that alter the direction and shape of the bunches in the collision zone
affect the overall luminosity. The most significant are: the hourglass effect, the collision
angle, and beam--beam interactions. The hourglass effect and the collision angle can be
treated analytically in isolation, but since they are interrelated, numerical calculation is
required for the highest accuracy.

\subsection{Hourglass effect}

\begin{figure}[htbp]
\centering
\includegraphics[width=0.5\linewidth]{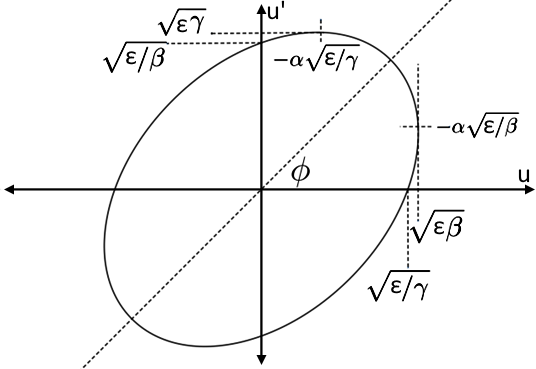}
\caption{The phase space, the ellipse formed by the particle distribution, and the special
points that determine its shape in terms of Twiss parameters.}
\label{fig:phasespace}
\end{figure}

The particle beams are focused by magnets towards the interaction point, so the transverse
sizes of the bunches change continually along the $s$-axis; this is the \emph{hourglass
effect}. The phase space is described by the equation of an ellipse enclosing the particles,
\begin{equation}
\varepsilon = \gamma u^{2} + 2\alpha u u' + \beta u'^{2},
\end{equation}
where $u$ is the position axis and $u'=du/ds$ the slope axis (the transverse velocity in units of $c$); $\alpha$, $\beta$, $\gamma$ are
the Courant--Snyder (Twiss) parameters~\cite{courantsnyder} and $\varepsilon$ the emittance. The ellipse area
$\pi\varepsilon$ remains constant along the path (Fig.~\ref{fig:phasespace}).

\begin{figure}[htbp]
\centering
\includegraphics[width=0.45\linewidth]{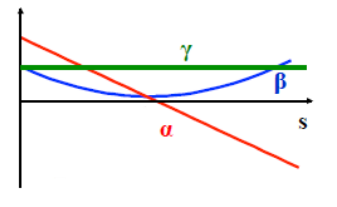}
\caption{The Twiss parameters of the beam as it moves along the $s$-axis without the
influence of the focusing-magnet field (the emittance is invariant).}
\label{fig:twiss}
\end{figure}

The connection between the $\beta$ function and the beam sizes is
\begin{equation}
\sigma_{t} = \sqrt{\varepsilon_{t}\beta_{t}}, \qquad t = x,y .
\end{equation}
The location where $\beta$ reaches its minimum value $\beta^{*}$ is the collision point,
with the smallest transverse sizes $\sigma^{*}$ (Fig.~\ref{fig:twiss}). The change in the transverse beam sizes
along $s$ is
\begin{equation}
\sigma_{t}^{2} = \sigma_{t}^{*2}\left( 1 + \frac{s^{2}}{\beta_{t}^{*2}} \right),
\qquad t = x,y .
\end{equation}
The hourglass effect thus alters the bunch shape during a collision, changing the luminosity.

\subsection{Collision angle}

\begin{figure}[htbp]
\centering
\includegraphics[width=0.5\linewidth]{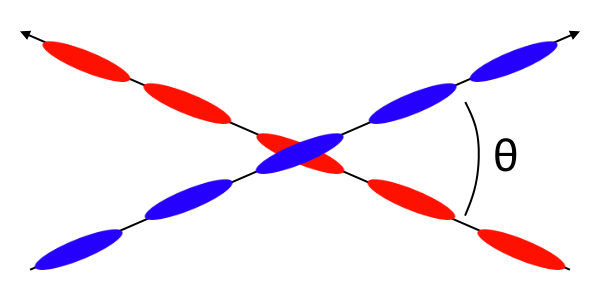}
\caption{Schematic of the collision of bunches at an angle. In reality the spacing between
bunches greatly exceeds the bunch length, so the angle is far smaller than depicted.}
\label{fig:angle}
\end{figure}

In colliders the beams are intentionally collided at an angle to prevent unintended
collisions outside the interaction region. For coasting (unbunched) beams the required angle
is large---at the Intersecting Storage Rings (ISR), which performed the first proton--proton
collisions in 1971, it was about $14.8^{\circ}$~\cite{isr}, causing a substantial luminosity reduction.
Bunched beams allow much smaller crossing angles; in the LHC the angle is roughly
$285~\mu\mathrm{rad}$~\cite{lhc} (Fig.~\ref{fig:angle}).

The impact of the collision angle on the luminosity of symmetric colliders~\cite{herrmuratori,piwinski,ruggiero} is
\begin{equation}
\mathcal{L} = \mathcal{L}_{0}\,S, \qquad
S = \frac{1}{\sqrt{1+\Phi^{2}}}, \qquad
\Phi = \frac{\sigma_{z}}{\sigma_{y}}\tan\theta ,
\end{equation}
where $\theta$ is the \emph{half} crossing angle (full angle $2\theta$); $\sigma_{y}$
appears because AloHEP implements the crossing in the longitudinal--vertical $(z,y)$
plane (Section~\ref{sec:numeric}), and for round beams $\sigma_{x}=\sigma_{y}$ so the
choice of plane is immaterial. For the LHC, with $2\theta = 285~\mu\mathrm{rad}$,
$\sigma_{x,y}=16.7~\mu\mathrm{m}$ and $\sigma_{s}=7.7~\mathrm{cm}$ (the example of Ref.~\cite{herrmuratori}), $S=0.835$, so the
luminosity decreases from $1.20\times10^{34}$ to $1.00\times10^{34}~\mathrm{cm^{-2}s^{-1}}$.

\subsection{Beam--beam interactions}

Particles within a bunch carry identical charges and tend to disperse, but at high speeds
this effect is suppressed by a factor inversely proportional to the square of the Lorentz
factor, so the beam can be compressed into very compact regions. This does not apply when
particles encounter particles of the oncoming bunch: opposite charges converge the particles
towards the centre, while like charges scatter them. This is the \emph{beam--beam
interaction}. The acceleration of a single particle on the $x$ axis is
\begin{equation}\label{eq:bbforce}
\frac{d^{2}x_{1}}{dt^{2}} = -\frac{4r_{1}N_{2}}{\gamma_{1}}
\int\!\!\int \frac{x_{1} - x}{\left( x_{1} - x \right)^{2} + \left( y_{1} - y \right)^{2}}\,
\rho_{2}\!\left( x,y,s,s_{0} \right)\,dx\,dy ,
\end{equation}
with an analogous expression for the $y$ axis. Equation~\eqref{eq:bbforce} is written for
oppositely charged bunches, the minus sign yielding attraction; for like charges the overall
sign is reversed, following the sign of the charge product of the two beams. For rigid
Gaussian bunches the two-dimensional integral admits the well-known closed form of
Bassetti and Erskine~\cite{bassetterskine}; once the bunches deform dynamically during the
crossing, however, the closed form no longer applies and the forces must be obtained
numerically. For light particles (electrons, positrons),
the transverse acceleration causes energy loss by photon emission (beamstrahlung~\cite{yokoyachen}), which
affects both the centre-of-mass energy and the luminosity. The combined influence of the
beam--beam interaction and the crossing angle has been studied extensively for $e^{+}e^{-}$
colliders~\cite{ohmi}; AloHEP makes both effects available for different species pairings.

%================================================================
\section{Numerical methods for the simulation of the collision region}\label{sec:numeric}

To reduce computational cost, some approximations are made. Since beams travel at extremely
high speeds (e.g.\ $0.999999991\,c$ for LHC protons) and the crossing angle is small, we
assume the colliding beams travel at the speed of light in opposite directions along the
$z$-axis; the three-dimensional motion is thus reduced to two-dimensional slices.
Furthermore, only the collision of two bunches need be simulated, with the total luminosity
obtained by multiplying by the bunch collision rate.

\begin{figure}[htbp]
\centering
\includegraphics[width=0.5\linewidth]{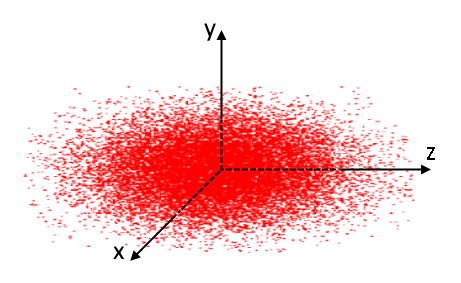}
\caption{Representative macroparticle distribution in a bunch, sampled from a normal
distribution with standard deviations $\sigma_{x}$, $\sigma_{y}$ and $\sigma_{z}$ on the
$x$, $y$ and $z$ axes respectively.}
\label{fig:mpdist}
\end{figure}

Because each bunch may contain $\sim\!10^{11}$ particles, we represent millions of real
particles by a single \emph{macroparticle}; each macroparticle has its own transverse
position and velocity, but moves at the speed of light along $z$. Bunches are thus modelled
as collections of tens of thousands of macroparticles (Fig.~\ref{fig:mpdist}).

\begin{figure}[htbp]
\centering
\includegraphics[width=0.65\linewidth]{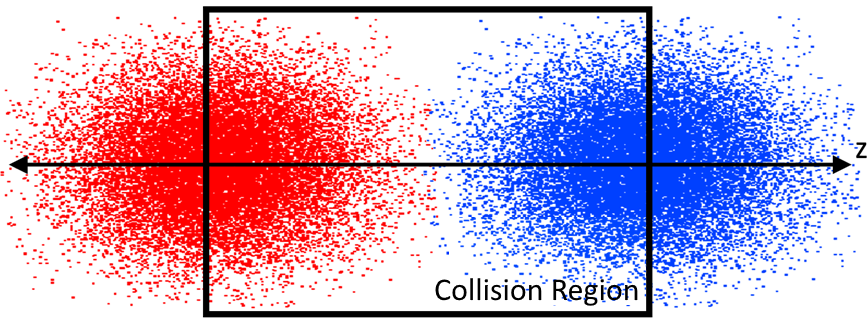}
\caption{Schematic of the collision region for symmetric, head-on colliding bunches. The
initial positions are set so the bunches intersect only within the collision region.}
\label{fig:region}
\end{figure}

The collision region is defined from the transverse and longitudinal sizes of the crossing
bunches:
\begin{equation}
\begin{aligned}
S_{x,y} &= k \times \max\!\left[ \sigma_{x_{1},y_{1}},\, \sigma_{x_{2},y_{2}} \right], \\
S_{z}   &= k \times \left( \sigma_{z_{1}} + \sigma_{z_{2}} \right)/2 ,
\end{aligned}
\end{equation}
where $S$ is the collision-region half-size---the simulated region spans $\pm S$ about the
collision point along each axis---and $k$ a coefficient chosen from the normal
distribution (e.g.\ $k=3$ includes $99.7\%$ of the particles, Fig.~\ref{fig:region}; all
benchmarks of Section~\ref{sec:benchmarks} use the default $k=4$, and the Gaussian sampling of
the macroparticles is truncated at $\pm k\sigma$ in position and angle so that every
macroparticle starts inside the simulated region). When
the hourglass effect is enabled and $\sigma_z/\beta_y^{*}>1/2$, the vertical half-size is
additionally multiplied by $2\sigma_z/\beta_y^{*}$, so that the diverging beam remains inside
the grid away from the waist.

\begin{figure}[htbp]
\centering
\includegraphics[width=0.8\linewidth]{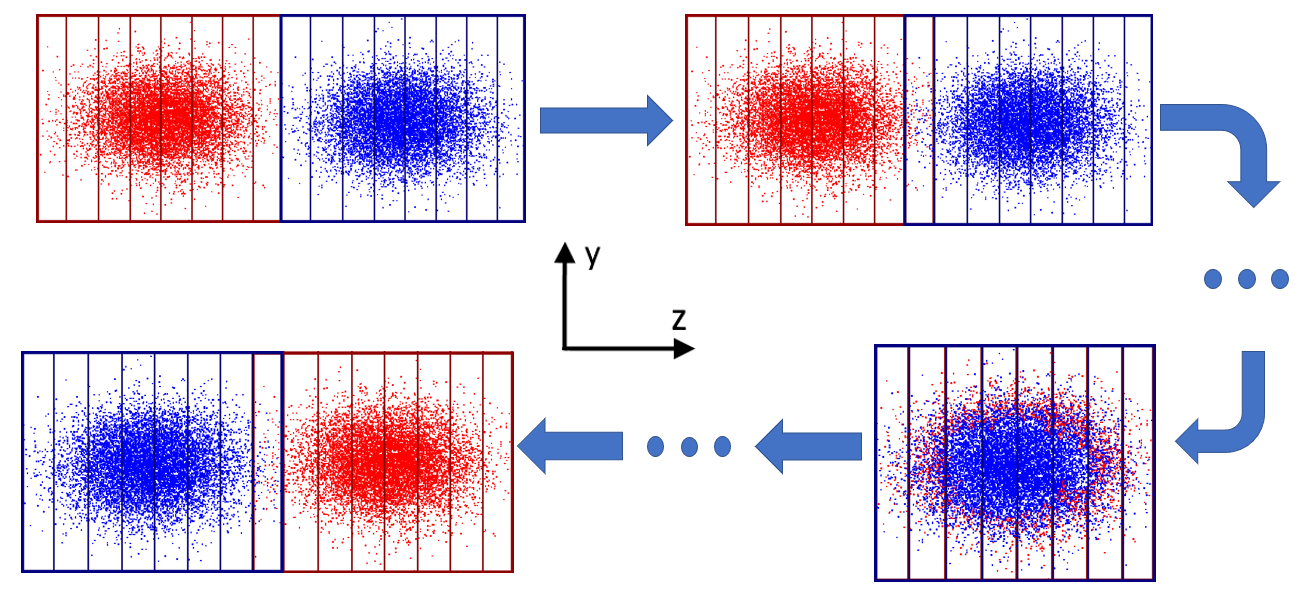}
\caption{Slice-by-slice movement of the crossing bunches, reducing the three-dimensional
distribution in each slice to two dimensions on the $x$ and $y$ axes.}
\label{fig:slices}
\end{figure}

Assuming the bunches travel at $c$ along $z$, each is divided into slices of length
$\Delta z = 2c\,\Delta t$ (the factor of two because the bunches approach each other at the relative speed $2c$). The
intersection densities of overlapping slices are computed and the slices stepped along $z$
(Fig.~\ref{fig:slices}).

\begin{figure}[htbp]
\centering
\includegraphics[width=0.35\linewidth]{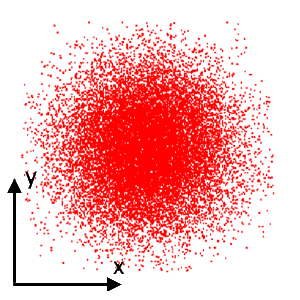}\hfill
\includegraphics[width=0.35\linewidth]{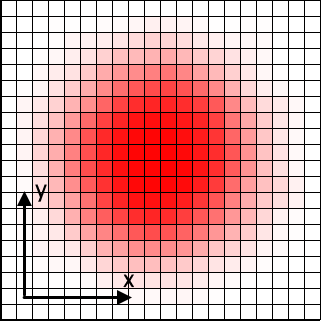}
\caption{A bunch slice reduced to two dimensions (left) and the distribution of its
macroparticles into cells (right).}
\label{fig:cells}
\end{figure}

To resolve each slice with fewer macroparticles, the slices are divided into rectangular
cells of size $\Delta x \times \Delta y$; the number of cells sets the transverse resolution
(Fig.~\ref{fig:cells}).

\begin{figure}[htbp]
\centering
\includegraphics[width=0.4\linewidth]{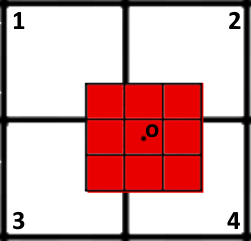}
\caption{Distribution of one macroparticle into cells. The shaded rectangle is a
macroparticle centred at $O$; the numbered rectangles are cells. For a macroparticle of nine
billion particles, one billion fall in cell~1, two billion each in cells~2 and~3, and four
billion in cell~4.}
\label{fig:cloud}
\end{figure}

Treating macroparticles as point-like produces sharp cell-to-cell transitions. This is
avoided with the \emph{particle cloud} (cloud-in-cell) method~\cite{birdsall} common in plasma physics: each macroparticle is treated as a uniformly filled rectangle, and its real particles are
distributed into the cells in proportion to the overlap areas (Fig.~\ref{fig:cloud}).

After each cloud-in-cell deposition (performed at every time step for each pair of
overlapping slices of the two bunches), the intersection with the corresponding slice of the opposing bunch is added to
the luminosity:
\begin{equation}\label{eq:lumnum}
\mathcal{L} = 2N_{1}N_{2}f_{c}\,
\frac{c\,\Delta t}{\Delta x\,\Delta y\,\Delta z\,N_{p}^{2}}
\sum_{n}\sum_{k}\sum_{ij} Q_{ij}^{(1)}Q_{ij}^{(2)} ,
\end{equation}
where $N_{1}$, $N_{2}$ are the bunch populations, $f_{c}$ the bunch collision rate, $N_{p}$
the number of macroparticles, $n$ each time step, $k$ each slice, and $Q_{ij}^{(1)}$,
$Q_{ij}^{(2)}$ the particle distributions of the intersecting slices. A prefactor of
$10^{-4}$ converts the result from $\mathrm{m^{-2}\,s^{-1}}$ to
$\mathrm{cm^{-2}\,s^{-1}}$. The factors affecting the luminosity are introduced in the
subsections below.

\subsection{Simulation of the hourglass effect}

\begin{figure}[htbp]
\centering
\includegraphics[width=0.7\linewidth]{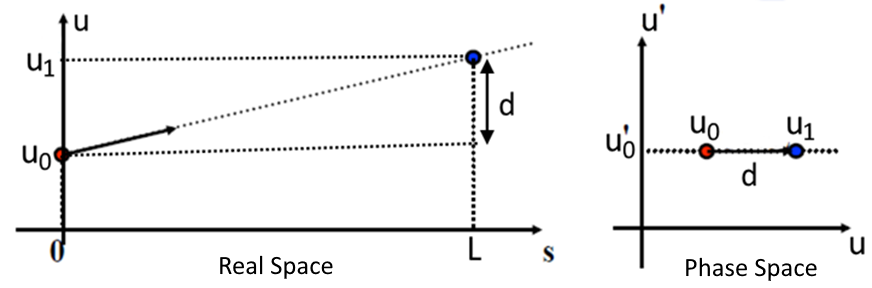}
\caption{Real-space and phase-space graphs for a particle in a bunch travelling in vacuum
without a magnetic field.}
\label{fig:vacuum}
\end{figure}

To simulate the hourglass effect, the motion of the particles leaving the magnets in vacuum
is applied to the macroparticles. Outside the magnetic field a particle maintains the
velocity at which it left the magnet, so its motion in real and phase space is as shown in
Fig.~\ref{fig:vacuum}.

\begin{figure}[htbp]
\centering
\includegraphics[width=0.85\linewidth]{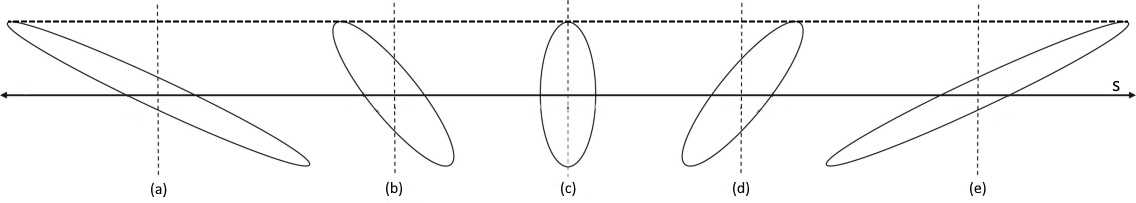}
\caption{Phase space of a bunch moving in field-free vacuum: shape (a) just after leaving
the field, (c) at the collision point, and (e) after dispersing beyond the collision point.}
\label{fig:phasechange}
\end{figure}

\begin{figure}[htbp]
\centering
\includegraphics[width=0.4\linewidth]{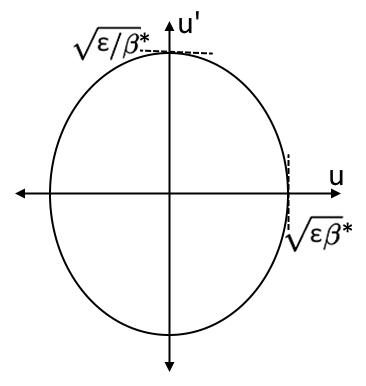}
\caption{Phase space at the collision point with the parameters expressed in terms of Twiss
parameters.}
\label{fig:phasetwiss}
\end{figure}

The motion in field-free vacuum is
\begin{equation}
\begin{aligned}
u_{0}' &= u_{1}', \\
u &= u_{0} + u_{0}' \cdot L ,
\end{aligned}
\end{equation}
where $L$ is the distance travelled along $s$. Since velocities are constant, defining the
initial positions and velocities suffices to add the hourglass effect. The phase space of
the focused beam at the collision point is shown in Figs.~\ref{fig:phasechange}
and~\ref{fig:phasetwiss}.

The phase-space boundaries on the $u$-axis give the transverse beam sizes
$\sigma_{x,y}^{*} = \sqrt{\varepsilon_{x,y}\beta_{x,y}^{*}}$ at the collision point. The
velocity distribution there is normal with standard deviation $\sqrt{\varepsilon/\beta^{*}}$;
since the velocities are invariant in vacuum, ${u'}^{*} = u_{0}'$. With the longitudinal
positions distributed normally and the
projected distance $L$ to the collision point known, the initial positions are
\begin{equation}
u_{0} = u^{*} - L \cdot u_{0}' .
\end{equation}

\subsection{Simulation of the crossing angle}

A finite crossing angle is implemented as a rigid rotation of each bunch in the
longitudinal--vertical $(z,y)$ plane, i.e.\ about the horizontal $x$-axis, applied before the
collision loop; the transverse horizontal coordinate $x$ is unchanged. The crossing-angle
parameter $\theta$ is the half crossing angle: the two counter-propagating bunches are rotated
by $-\theta$ and $+\theta$, so that the full angle between the beam axes is $2\theta$ and the
beams cross in the vertical plane. For a particle at $(x_0,y_0,z_0)$,
\begin{equation}
\begin{pmatrix} z_\theta \\ y_\theta \end{pmatrix}
=
\begin{pmatrix} \cos\theta & -\sin\theta \\ \sin\theta & \cos\theta \end{pmatrix}
\begin{pmatrix} z_0 \\ y_0 \end{pmatrix},
\qquad x_\theta = x_0 ,
\label{eq:crossing}
\end{equation}
and each particle receives the corresponding vertical velocity component (evaluated in the
ultra-relativistic limit $|v|\approx c$). The resulting geometric-overlap reduction is governed by the Piwinski factor~\cite{piwinski,ruggiero} $S=1/\sqrt{1+\Phi^{2}}$ of Section~\ref{sec:analysis} (with $\theta$ the half angle of the $(z,y)$-plane rotation applied here).

\subsection{Simulation of beam--beam interactions}

\begin{figure}[htbp]
\centering
\includegraphics[width=0.5\linewidth]{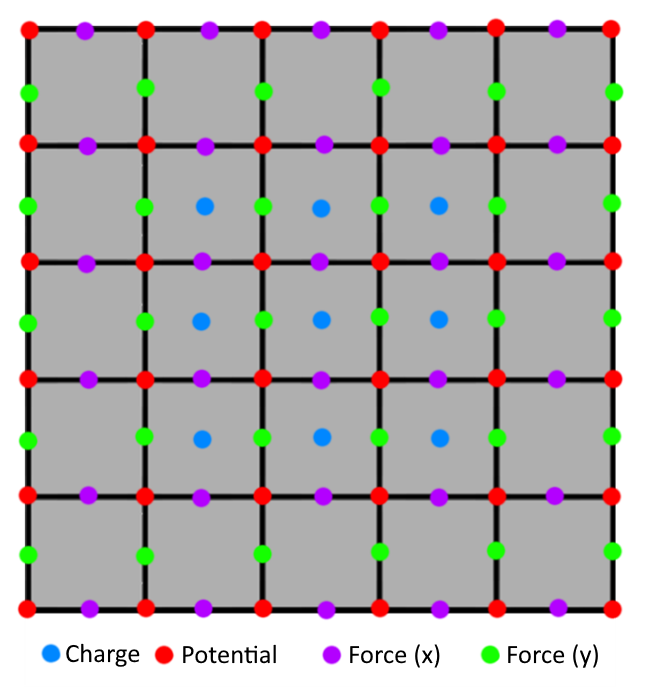}
\caption{Charge, potential and force matrix elements on a slice of a bunch divided into
$5\times5$ cells.}
\label{fig:matrices}
\end{figure}

During the collision the bunches deform under each other's electromagnetic field, affecting
the luminosity. Because of the relativistic motion the fields are transverse, and since the
bunch length $\sigma_{z}$ greatly exceeds $\sigma_{x},\sigma_{y}$, only overlapping slices
interact at each step. The cell occupancies represent the charge distribution, from which
the slice potential at any $(x,y)$ is
\begin{equation}
\phi(x,y) = \sum_{i,j} Q_{ij}\,
\ln\!\left[ (x - i\Delta x)^{2} + (y - j\Delta y)^{2} \right]/2 ,
\end{equation}
with $Q_{ij}$ the charge-matrix elements. Taking the charges at the cell centres and the
potential at the corners $(m,n)$,
\begin{equation}
\begin{aligned}
T_{i,j} &= \ln\!\left( (i - \tfrac12)^{2}\Delta x^{2} + (j - \tfrac12)^{2}\Delta y^{2} \right)/2 , \\
\phi_{mn} &= \phi(m\Delta x, n\Delta y) = \sum_{i,j} T_{m-i,\,n-j}\,Q_{ij} .
\end{aligned}
\end{equation}
Evaluating the potential at the corners (rather than centres) removes the singularity that
would otherwise arise where the charge--potential distance vanishes. The forces follow from
finite differences:
\begin{equation}
\begin{aligned}
F_{x}\!\left( (m + \tfrac12)\Delta x, n\Delta y \right) &= \left( \phi_{m+1,n} - \phi_{mn} \right)/\Delta x , \\
F_{y}\!\left( m\Delta x, (n + \tfrac12)\Delta y \right) &= \left( \phi_{m,n+1} - \phi_{mn} \right)/\Delta y .
\end{aligned}
\end{equation}
The direct sum yields the free-space (open-boundary) potential, so no boundary condition
is imposed; the potential grid extends two cells beyond the charge grid so that the
interpolation stencil of the force (below) is complete everywhere inside the collision
region. An example at $5\times5$ resolution is given in
Fig.~\ref{fig:matrices}.

\begin{figure}[htbp]
\centering
\includegraphics[width=0.65\linewidth]{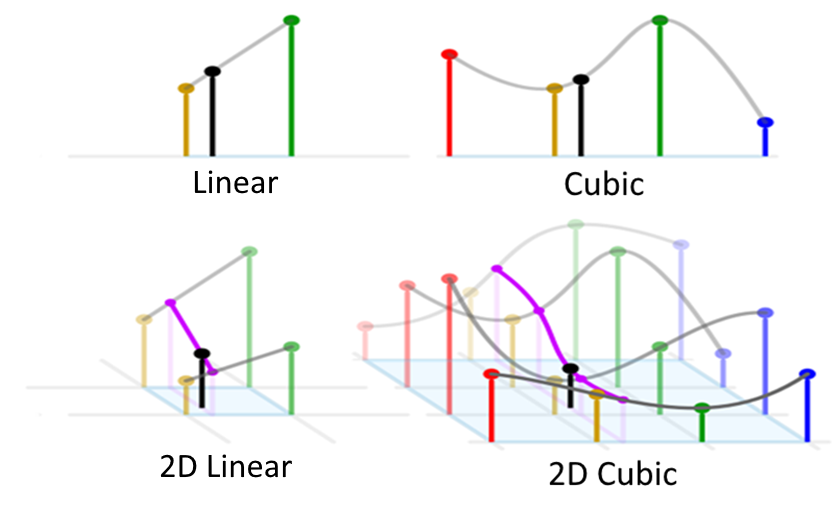}
\caption{Schematic of one- and two-dimensional linear and cubic interpolation.}
\label{fig:interp}
\end{figure}

The forces acting on each macroparticle are interpolated from the force grid. Bilinear
interpolation over the four neighbouring nodes is only $C^{0}$: the force is continuous, but
its gradient jumps at every cell boundary, so the cell structure can imprint itself on a
strongly disrupted beam. To avoid this, AloHEP uses a $C^{1}$
two-dimensional bicubic (Catmull--Rom) interpolation over the $16$ neighbouring nodes
(Fig.~\ref{fig:interp}).

Once the forces are obtained, the beam--beam acceleration of a macroparticle is
\begin{equation}\label{eq:pinchacc}
a_{x}^{n} = -\frac{4r_{1}N_{2}}{\gamma_{1}^{n}N_{p}\Delta z}\,
F_{x}^{(2)}\!\left( x_{n}, y_{n} \right),
\end{equation}
where $r_{1}$ is the classical radius of the (accelerated) beam-1 particle, $\gamma_{1}^{n}$
the Lorentz factor at step $n$, and $F_{x}^{(2)}$ the interpolated force; an analogous
relation holds for the $y$ axis. As in Eq.~\eqref{eq:bbforce}, the minus sign corresponds to
oppositely charged beams: the code applies the overall sign $-\operatorname{sgn}(q_{1}q_{2})$,
so same-charge beams are defocused (cf.\ the $\pm$ in Fig.~\ref{fig:flowchart} and the sign
verification of Section~\ref{sec:pinch}). Repeating this for every macroparticle in each intersecting
slice and time step extends the simulation to include beam--beam interactions.

%================================================================
\section{Software architecture}\label{sec:software}

AloHEP is implemented entirely in Java (Java~7-compatible syntax; it builds and runs
unchanged on OpenJDK~8--21, and all benchmarks were run with OpenJDK~17) and
runs on any platform with a JVM; its only external dependency is Gson~2.9.1~\cite{gson} for JSON
handling. The design is object-oriented and separates the graphical front end from the
computational engine, so the engine can run either interactively or in a fully headless
(command-line) mode.

A central \texttt{DataManager} parses pre-configured accelerator parameter sets
from a JSON tree (\texttt{save/}) at runtime, grouped by particle type and tagged
\textsc{linear} or \textsc{circle} according to the name of the particle folder
(e.g.\ \texttt{electron-linac}); this tag selects whether the results report disruption
parameters $D_{x,y}$ (linear machines) or beam--beam tune shifts $\xi_{x,y}$ (circular
machines). Because the database is data-driven, an entirely novel asymmetric scenario---for
example an electron beam against a heavy-ion beam---can be defined and simulated without
recompiling.

The engine mirrors the physical decomposition of the problem:
\begin{itemize}
\item \texttt{MacroParticle}: a statistically significant cluster of real particles carrying
its 3D position, transverse velocity and acceleration, and Lorentz factor $\gamma$.
\item \texttt{Slice}: the Particle-in-Cell domain, holding the charge matrix $Q_{ij}$, the
potential $\phi_{mn}$ and the force matrices $F_{x}$, $F_{y}$, and performing the
cloud-in-cell charge assignment, the Green's-function potential solve, the finite-difference
forces, the bicubic interpolation, and the slice-overlap luminosity integral
$\sum_{ij}Q_{ij}^{(1)}Q_{ij}^{(2)}$.
\item \texttt{Beam} and \texttt{Bunch}: the per-beam layer. \texttt{Beam} pairs a particle
species with an accelerator parameter set from the database; \texttt{Bunch} derives the beam
sizes and Lorentz factor from it, samples the macroparticles from the Gaussian phase-space
distribution (with the hourglass modulation $x(z)=x_{0}-x'|z|$), applies the crossing-angle
rotation, distributes the macroparticles over the slices and advances them along $z$.

\item \texttt{LuminosityCalc}: the master controller driving the collision loop, accumulating
the nominal (analytic) and effective (simulated) luminosities, and deriving the disruption
parameters and beam--beam tune shifts.
\end{itemize}

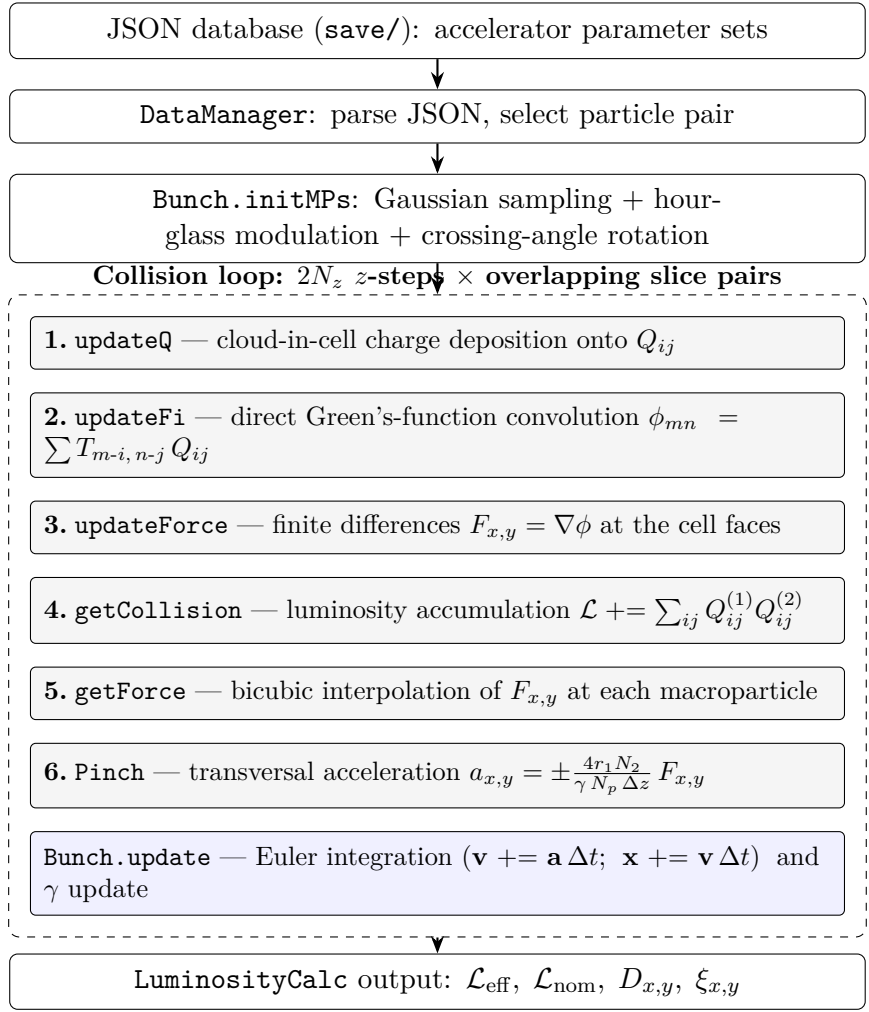
\begin{figure}[htbp]
\centering
\begin{tikzpicture}[
  >=Stealth, font=\small,
  box/.style={draw, rounded corners=3pt, align=center, inner sep=5pt,
              text width=0.80\linewidth, fill=white},
  stepbox/.style={draw, rounded corners=2pt, align=left, inner sep=5pt,
               text width=0.76\linewidth, fill=gray!8, font=\footnotesize},
  arr/.style={->, thick},
]

% --- Pipeline (outside loop) ---
\node[box] (json) {JSON database (\texttt{save/}): accelerator parameter sets};
\node[box, below=0.4cm of json] (dm) {\texttt{DataManager}: parse JSON, select particle pair};
\node[box, below=0.4cm of dm] (init) {\texttt{Bunch.initMPs}: Gaussian sampling + hourglass
  modulation + crossing-angle rotation};

% --- Collision loop steps ---
\node[stepbox, below=0.7cm of init] (s1)
  {\textbf{1.}\;\texttt{updateQ} --- cloud-in-cell charge deposition onto $Q_{ij}$};
\node[stepbox, below=0.3cm of s1] (s2)
  {\textbf{2.}\;\texttt{updateFi} --- direct Green's-function convolution
   $\phi_{mn}=\sum T_{m\text{-}i,\,n\text{-}j}\,Q_{ij}$};
\node[stepbox, below=0.3cm of s2] (s3)
  {\textbf{3.}\;\texttt{updateForce} --- finite differences $F_{x,y}=\nabla\phi$ at the cell faces};
\node[stepbox, below=0.3cm of s3] (s4)
  {\textbf{4.}\;\texttt{getCollision} --- luminosity accumulation
   $\mathcal{L}\mathrel{+}=\textstyle\sum_{ij} Q^{(1)}_{ij}Q^{(2)}_{ij}$};
\node[stepbox, below=0.3cm of s4] (s5)
  {\textbf{5.}\;\texttt{getForce} --- bicubic interpolation of $F_{x,y}$
   at each macroparticle};
\node[stepbox, below=0.3cm of s5] (s6)
  {\textbf{6.}\;\texttt{Pinch} --- transversal acceleration
   $a_{x,y}=\pm\frac{4r_1N_2}{\gamma\,N_p\,\Delta z}\,F_{x,y}$};

% --- Bunch update (inside loop, after all slices) ---
\node[stepbox, below=0.3cm of s6, fill=blue!6] (upd)
  {\texttt{Bunch.update} --- Euler integration
   ($\mathbf{v}\mathrel{+}=\mathbf{a}\,\Delta t$;\;
   $\mathbf{x}\mathrel{+}=\mathbf{v}\,\Delta t$)
   \;and\; $\gamma$ update};

% --- Output ---
\node[box, below=0.5cm of upd] (out)
  {\texttt{LuminosityCalc} output:
   $\mathcal{L}_{\rm eff}$,\;\,$\mathcal{L}_{\rm nom}$,\;\,$D_{x,y}$,\;\,$\xi_{x,y}$};

% --- Dashed loop frame ---
\node[draw, dashed, rounded corners=4pt, inner sep=8pt,
      fit=(s1)(s2)(s3)(s4)(s5)(s6)(upd),
      label={[font=\footnotesize\bfseries,yshift=-2pt]above:%
        Collision loop: $2N_z$ $z$-steps $\times$ overlapping slice pairs}] (loop) {};

% --- Arrows ---
\draw[arr] (json) -- (dm);
\draw[arr] (dm) -- (init);
\draw[arr] (init) -- (loop.north);
\draw[arr] (loop.south) -- (out);

\end{tikzpicture}
\caption{AloHEP simulation pipeline. The dashed box encloses the collision loop: steps~1--6
are executed for every pair of overlapping slices at each $z$-step, after which
\texttt{Bunch.update} advances all macroparticles (Euler integration, with optional beamstrahlung).
The loop iterates over $2N_z$ longitudinal steps; the final output is computed once by
\texttt{LuminosityCalc} after the bunches have fully crossed.}
\label{fig:flowchart}
\end{figure}

Figure~\ref{fig:flowchart} summarises the simulation pipeline from JSON input through the
collision loop to the final luminosity output.

A \emph{Matched Beams} helper equalises the transverse spot sizes by enlarging the
smaller-spot beam to the larger of the two, recomputing the $\beta$-functions from
$\beta = \sigma_{\max}^{2}/\varepsilon$. AloHEP provides two entry points: the Swing-based
GUI (\texttt{Launcher}), with real-time macroparticle animation, and the command-line
\texttt{HeadlessLauncher} for batch studies, e.g.

\begin{quote}\footnotesize
\texttt{java -cp ".:gson-2.9.1.jar:bin" \textbackslash}\\
\texttt{~~org.hepforge.alohep.HeadlessLauncher \textbackslash}\\
\texttt{~~-left proton-ele/EIC-PDG1 -right electron-ring/EIC-PDG \textbackslash}\\
\texttt{~~-hourglass -pinch -macrop 50000 \textbackslash}\\
\texttt{~~-xres 50 -yres 50 -zres 50 -csv}
\end{quote}

Both entry points are launched from the compiled classes with Gson on the classpath, e.g.
\begin{quote}\footnotesize
\texttt{java -cp ".:gson-2.9.1.jar:bin" org.hepforge.alohep.Launcher}
\end{quote}
for the GUI (substitute \texttt{HeadlessLauncher} for batch use). AloHEP is released under the MIT License and is available at \url{https://github.com/tobbetuyef/AloHEP}.

\subsection{Interactive use}\label{sec:gui}

\begin{figure}[htbp]
\centering
\includegraphics[width=0.9\linewidth]{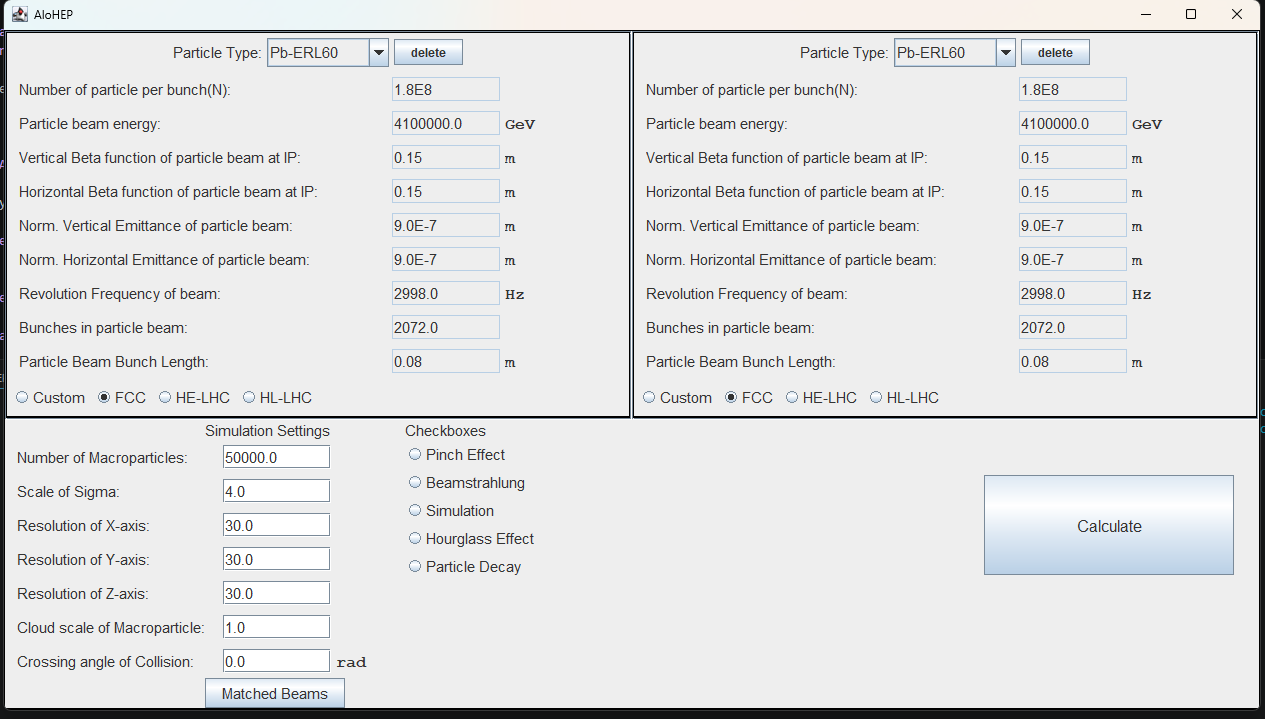}
\caption{The AloHEP graphical interface. The two panels across the top are the left and right
beam editors, each with a species drop-down, a machine selector (\texttt{Custom} for
user-defined beams) and the editable parameter fields. Below them, the \emph{Simulation
Settings} panel (left) holds the grid resolution, macroparticle count and the \emph{Matched
Beams} helper, the \emph{Checkboxes} panel (centre) toggles the individual physical effects,
and the \emph{Calculate} button (right) starts the run. Enabling the simulation additionally
opens a real-time macroparticle animation, and the final luminosity, reduction factor and
beam--beam figures of merit are shown in a separate results window.}
\label{fig:gui}
\end{figure}

The graphical interface gathers a complete simulation into a single window
(Fig.~\ref{fig:gui}). A typical session has three steps---define the two beams, choose which
physical effects to include, and press \emph{Calculate}---after which the results are
displayed in their own window. The panels of Fig.~\ref{fig:gui}, each identified by its
on-screen label, are described below; a full reference for every parameter (database keys,
units, and the rules by which the collision frequency and the storage time are derived from
them) is given in the repository \texttt{README} rather than repeated here.

\begin{enumerate}
\item \textbf{Beam editors.} The two panels across the top of the window configure the left
and right beams independently. For each, a drop-down selects the \emph{particle species} group and a row of
radio buttons selects a \emph{collider type} from the JSON database; choosing \texttt{Custom}
unlocks the parameter fields so the user can enter their own values---bunch population, beam
energy, $\beta^{*}_{x,y}$, normalised emittances, bunch length and revolution
frequency. A custom configuration can be named and stored back into the database with the
\emph{save} button, after which it reappears as a selectable machine; the same button removes
the currently selected entry (user-defined or shipped) from the database. This is how an entirely new or hypothetical collider is set up without
recompiling.
\item \textbf{Simulation settings.} In the \emph{Simulation Settings} panel (lower left) the
grid resolution ($N_x,N_y,N_z$), the macroparticle count, the collision-region coefficient
$k$ (labelled \emph{Scale of Sigma}) and the crossing angle (entered in radians) are set. The
\emph{Matched Beams} helper
equalises the transverse spot sizes (Section~\ref{sec:software}) for round-beam studies.
\item \textbf{Physical effects.} The \emph{Checkboxes} panel (centre) enables the hourglass
effect, the beam--beam pinch, beamstrahlung and particle decay; each can be toggled independently, with
the dependencies enforced automatically (enabling pinch, hourglass or beamstrahlung turns on
the macroparticle simulation, and beamstrahlung requires the pinch). With all boxes cleared
no macroparticle simulation is run: the \emph{Matched Beams} helper is applied and the
analytic nominal luminosity $\mathcal{L}_{\rm nom}$ is reported.
\item \textbf{Run.} The \emph{Calculate} button (right) starts the run (and becomes
\emph{Cancel} while it is in progress); the progress bar tracks the bunch crossing. When the simulation is enabled, a
separate window shows the macroparticles in real time, so bunch deformation can be watched as
it develops.
\end{enumerate}

On completion a \emph{results} window reports the centre-of-mass energy, the nominal and
effective luminosities and their ratio, together with the per-beam spot sizes $\sigma_{x,y}$
and---depending on whether the machine is tagged linear or circular---either the disruption
parameters $D_{x,y}$ or the beam--beam tune shifts $\xi_{x,y}$ (Section~\ref{sec:software}).
Each quantity is labelled, so the figure of merit the user is after can be read off directly.

%================================================================
\section{Benchmarks and verification}\label{sec:benchmarks}

All benchmarks were run with the headless interface on a single thread (serial garbage
collector, 4~GB heap limit via \texttt{-Xmx4g}), utilising a single core of an AMD Ryzen~9~9900X processor
(4.4~GHz base clock, twelve cores available), with 48~GB RAM under Windows~11, using
OpenJDK~17.0.19. AloHEP performs all computation on the CPU; no GPU acceleration is used. Each
configuration was executed three times (the macroparticle sampling uses an unseeded random
generator, so repeated runs differ statistically); the values reported below are the means of
the three repetitions. The run-to-run standard deviation of the luminosity is below $\sim\!1\%$
for the symmetric verification runs and below $\sim\!0.6\%$ for both HERA and EIC at
$5\times10^4$ macroparticles.

%----------------------------------------------------------------
\subsection{Verification against the analytic limit}\label{sec:verif}

\begin{figure}[H]
\centering
\includegraphics[width=0.75\linewidth]{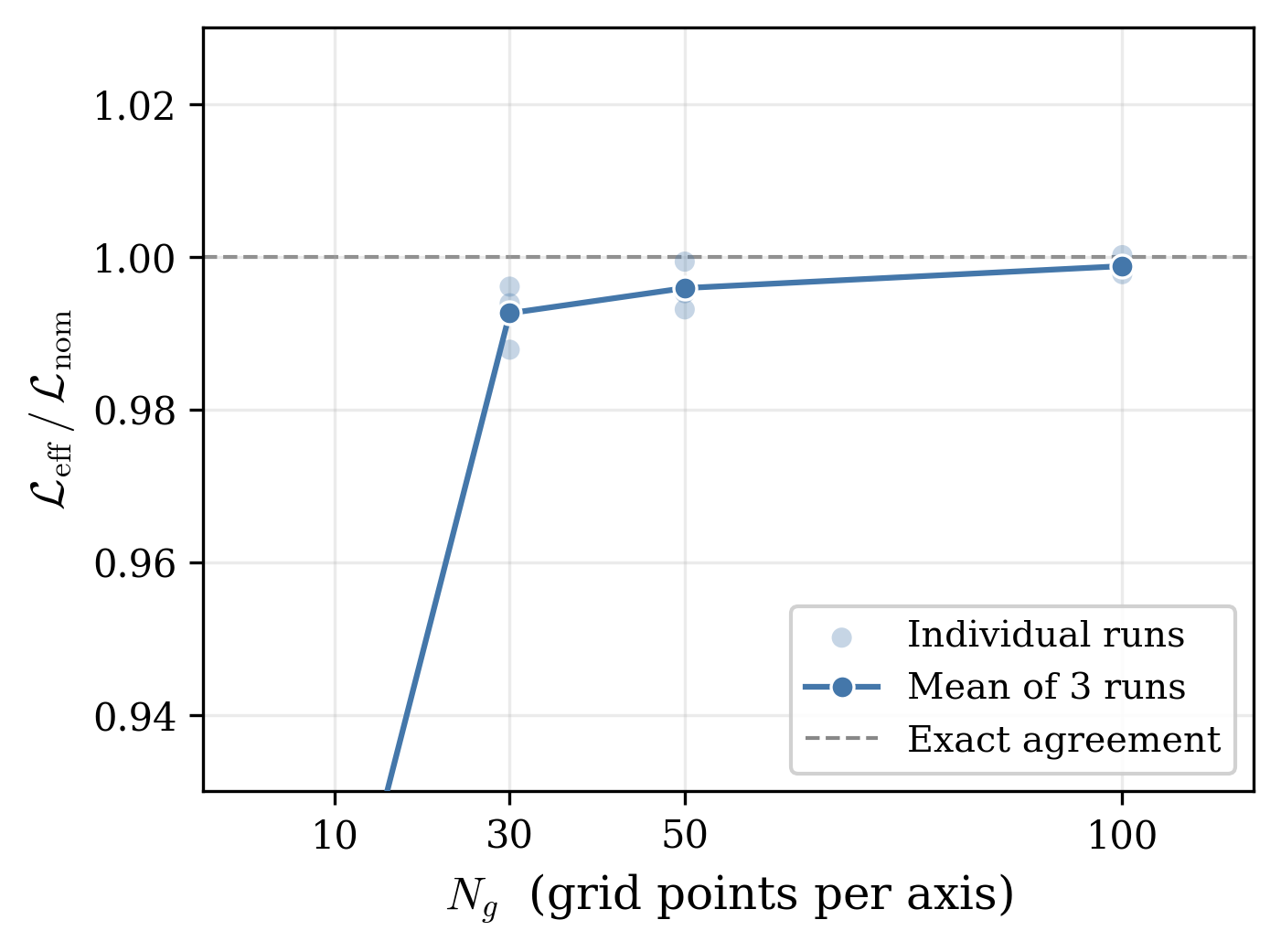}
\caption{Ratio of the simulated to the analytic geometric luminosity as a function of the
transverse grid resolution $N_g$, for a symmetric head-on $e^{+}e^{-}$ collision with no
perturbing effects. The dashed line marks exact agreement.}
\label{fig:conv}
\end{figure}

To verify the numerical engine we simulate a symmetric, head-on $e^{+}e^{-}$ collision
(ILC-125~\cite{ilctdr}, $\sqrt{s}=250$~GeV) with all perturbing effects disabled, so that the simulated
luminosity $\mathcal{L}_{\rm eff}$, accumulated via Eq.~\eqref{eq:lumnum}, must reproduce the analytic geometric luminosity
$\mathcal{L}_{\rm nom}$, which for these identical beams coincides with
$\mathcal{L}_0$ of Eq.~\eqref{eq:lumgeo0}. Table~\ref{tab:conv} and Fig.~\ref{fig:conv} show the
result as a function of the transverse grid resolution $N_g$, at a fixed $5\times10^{4}$
macroparticles. The coarsest grid ($10^3$) underestimates the luminosity by $\sim\!10\%$,
as expected from the cloud-in-cell smoothing of an under-resolved distribution: the
cloud-in-cell assignment convolves each gridded beam with a triangular kernel of variance
$\Delta^{2}/6$, so for a cell size $\Delta=2k\sigma/N_g$ the overlap integral is reduced by
$[1+(2k/N_g)^{2}/6]^{-1}$, i.e.\ by the factors $0.904$, $0.988$, $0.996$ and $0.999$ for
$N_g=10$, $30$, $50$ and $100$ with $k=4$, in agreement with Table~\ref{tab:conv}. As $N_g$
increases the ratio converges monotonically towards unity, reaching $0.993$ at $30^3$ (with a
run-to-run standard deviation of $0.4\%$; the same configuration run with $10^{6}$
macroparticles gives $0.988$, \ref{app:benchmap}) and
$\mathcal{L}_{\rm eff}/\mathcal{L}_{\rm nom}=0.999$ at $100^3$. The
engine therefore reproduces the analytic geometric luminosity at the sub-percent level once the
grid is adequately resolved.

\begin{table}[H]
\centering
\caption{Grid-convergence verification (ILC-125, $\sqrt{s}=250$~GeV, $5\times10^4$
macroparticles, no perturbing effects). $\mathcal{L}_{\rm nom}=3.786\times10^{33}\,
\mathrm{cm^{-2}s^{-1}}$.}
\label{tab:conv}
\begin{tabular}{lcccc}
\toprule
Grid $N_g^3$ & Runtime & $\mathcal{L}_{\rm eff}$ [$\mathrm{cm^{-2}s^{-1}}$] &
$\mathcal{L}_{\rm eff}/\mathcal{L}_{\rm nom}$ & Potential-solve share \\
\midrule
$10^3$   & $0.092$~s & $3.420\times10^{33}$ & $0.9032$ & $47\%$ \\
$30^3$   & $12.2$~s  & $3.758\times10^{33}$ & $0.9926$ & $97.5\%$ \\
$50^3$   & $208$~s   & $3.771\times10^{33}$ & $0.9959$ & $99.6\%$ \\
$100^3$  & $3.20$~h  & $3.782\times10^{33}$ & $0.9988$ & $\sim\!100\%$ \\
\bottomrule
\end{tabular}
\end{table}

%----------------------------------------------------------------
\subsection{Computational scaling}\label{sec:scaling}

\begin{figure}[htbp]
\centering
\includegraphics[width=1.00\linewidth]{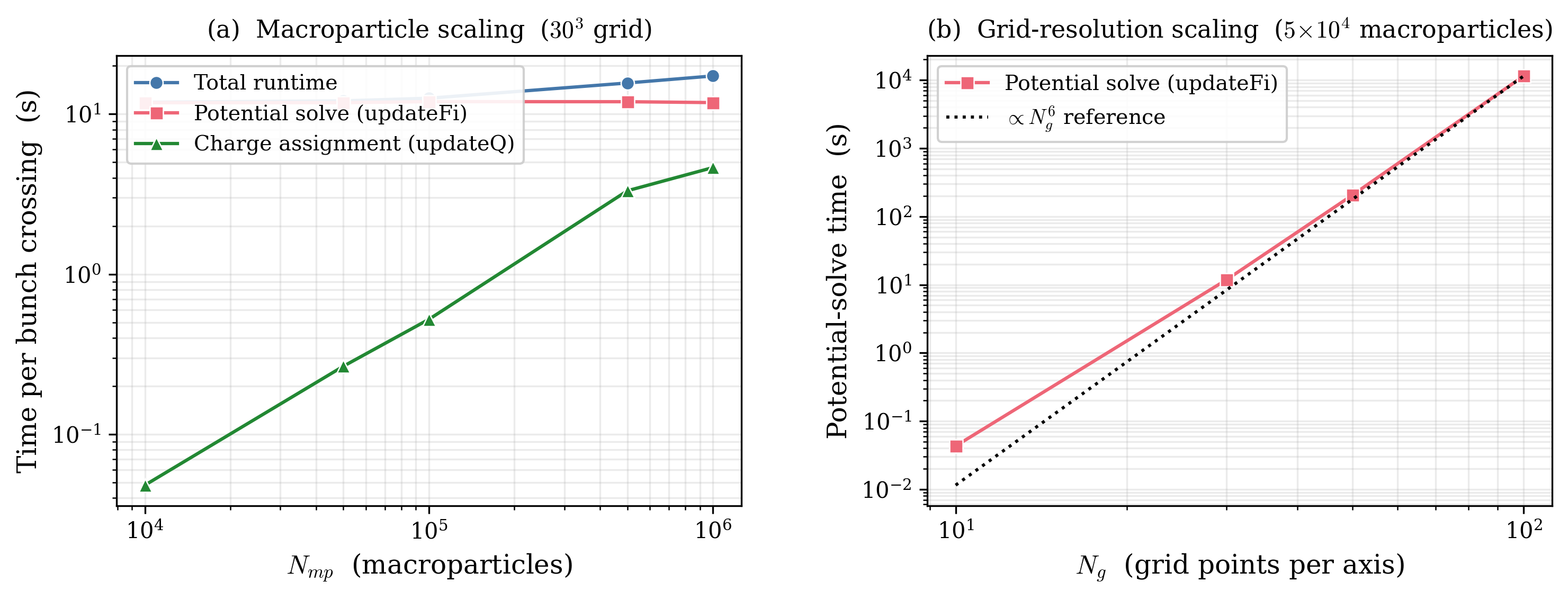}
\caption{Computational scaling of AloHEP. (a) Runtime per bunch crossing versus
macroparticle number at a fixed $30^3$ grid: the potential solve is constant while the
charge assignment grows linearly. (b) Potential-solve time versus grid resolution at
$5\times10^4$ macroparticles, with the $\propto N_g^6$ reference (dotted).}
\label{fig:scaling}
\end{figure}

The per-phase profiler isolates the cost of each stage of the collision loop.
Table~\ref{tab:npscale} and Fig.~\ref{fig:scaling}(a) show the runtime as a function of the
macroparticle number $N_{mp}$ at a fixed $30^3$ grid. The direct Green's-function potential
solve (\texttt{updateFi}) is independent of $N_{mp}$ at $\sim\!11.85$~s and dominates the
runtime up to $N_{mp}\sim10^5$ (the potential and force grids are computed at every step
whether or not the pinch is enabled, so pinch-off runs cost the same as pinch-on runs); only
the per-macroparticle stages, the charge assignment (\texttt{updateQ}) and the Euler push, grow
with the macroparticle count, and they do so approximately linearly (a fitted exponent of
$\approx1.0$ for \texttt{updateQ}, consistent with the $\mathcal{O}(N_{mp})$ cloud-in-cell deposition).
Consequently the total runtime rises only from $11.8$~s at $10^4$ macroparticles to $17.30$~s
at $10^6$.

Figure~\ref{fig:scaling}(b) shows the much steeper dependence on the grid resolution. The
potential solve scales as $\mathcal{O}(N_x^2 N_y^2)$ per slice and is repeated for every
overlapping slice pair at each of the $2N_z$ longitudinal steps, i.e.\ $\mathcal{O}(N_z^2)$
times per crossing, so the total cost scales theoretically as $\mathcal{O}(N_g^6)$ for a cubic
grid. The measured exponent rises from $5.1$ ($10\!\to\!30$) to $5.8$ ($50\!\to\!100$),
approaching the theoretical bound as the lower-order initialisation and bookkeeping terms
become negligible. 

\begin{table}[htbp]
\centering
\caption{Macroparticle scaling at a fixed $30^3$ grid (ILC-125, no perturbing effects).
Times are means of three runs.}
\label{tab:npscale}
\resizebox{\textwidth}{!}{%
\begin{tabular}{lccc}
\toprule
$N_{mp}$ & Total runtime & Charge assignment (\texttt{updateQ}) & Potential solve (\texttt{updateFi}) \\
\midrule
$1\times10^4$ & $11.84$~s & $48$~ms   & $11.78$~s \\
$5\times10^4$ & $12.09$~s & $266$~ms  & $11.78$~s \\
$1\times10^5$ & $12.56$~s & $523$~ms  & $11.96$~s \\
$5\times10^5$ & $15.63$~s & $3.32$~s  & $11.94$~s \\
$1\times10^6$ & $17.30$~s & $4.63$~s  & $11.79$~s \\
\bottomrule
\end{tabular}%
}
\end{table}

%----------------------------------------------------------------
\subsection{Hourglass effect}\label{sec:hourglassbench}

\begin{figure}[htbp]
\centering
\includegraphics[width=0.6\linewidth]{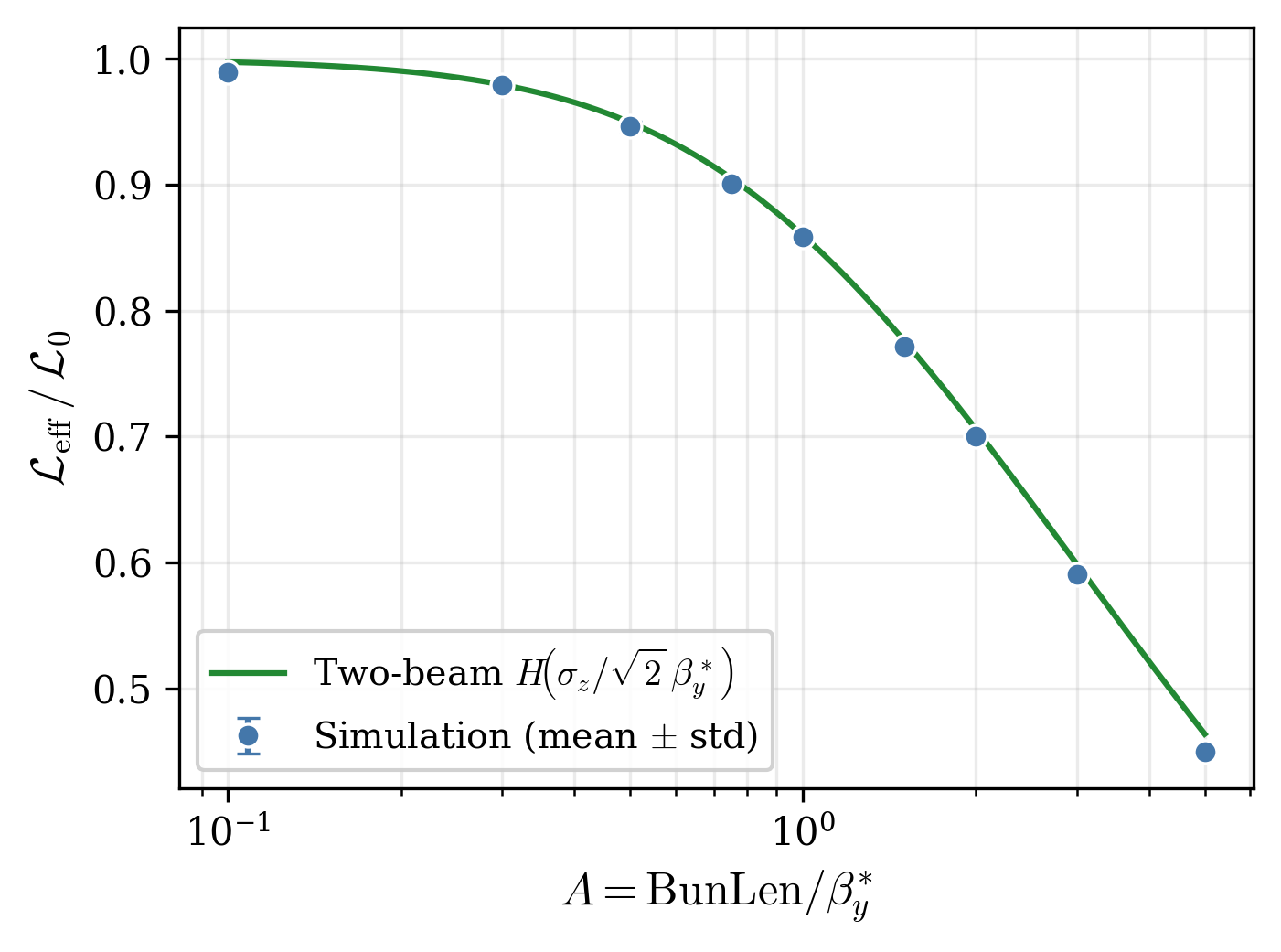}
\caption{Hourglass luminosity reduction for symmetric ILC $e^{+}e^{-}$ versus the formal
parameter $A=\mathrm{BunLen}/\beta_y^{*}$: simulation (points, mean of three runs) and the
two-beam analytic factor of Eq.~\eqref{eq:Hflat} (curve).}
\label{fig:hourglass}
\end{figure}

To verify the hourglass treatment of Sections~\ref{sec:analysis} and~\ref{sec:numeric} in
isolation, we collide symmetric ILC $e^{+}e^{-}$ bunches with the pinch and crossing angle
disabled and sweep the formal parameter $A=\mathrm{BunLen}/\beta_y^{*}$ by varying
$\beta_y^{*}$ (with $\mathrm{BunLen}$ fixed at the ILC value $0.3$~mm), on a
$50\times100\times50$ grid ($N_x\times N_y\times N_z$) with $5\times10^{4}$ macroparticles
(mean of three runs). The transverse grid is deliberately refined in the vertical plane:
here the focusing acts only vertically (the horizontal hourglass is negligible,
$A_x\simeq0.02$, so the horizontal beam size is essentially constant through the crossing),
and it is the shrinking vertical waist $\sigma_y^{*}$ that sets the resolution the simulation
must deliver. Because luminosity is a \emph{two-beam}
collision, a pair of particles at intra-bunch longitudinal positions $\xi_1$ and $\xi_2$ meets
at $s=(\xi_1+\xi_2)/2$, so the collision points are distributed with r.m.s.\ $\sigma_z/\sqrt2$;
the matching analytic reduction is therefore the flat-beam hourglass factor~\cite{yokoyachen,furman} evaluated at this
collision length,
\begin{equation}
H\!\left(\frac{\sigma_z}{\sqrt2\,\beta_y^{*}}\right)=
\frac{1}{\sqrt{2\pi}}\int_{-\infty}^{\infty}
\frac{e^{-u^{2}/2}\,du}
{\sqrt{1+\bigl(\tfrac{\sigma_z u}{\sqrt2\,\beta_x^{*}}\bigr)^{2}}\;
 \sqrt{1+\bigl(\tfrac{\sigma_z u}{\sqrt2\,\beta_y^{*}}\bigr)^{2}}},
\label{eq:Hflat}
\end{equation}
with $\sigma_z=\mathrm{BunLen}$.
Figure~\ref{fig:hourglass} shows that the simulated $\mathcal{L}_{\rm eff}/\mathcal{L}_{0}$
reproduces Eq.~\eqref{eq:Hflat} to within $\sim\!1\%$ across the full range $A\lesssim1.5$
that spans realistic machines (the ILC sits at $A\simeq0.75$), with deviations nowhere
exceeding $\sim\!0.8\%$; we independently confirmed Eq.~\eqref{eq:Hflat} against a direct two-beam
Monte-Carlo integration of the collision overlap. At larger $A$ the simulated points fall
slightly below the analytic curve---about $0.7\%$ at $A=2$, $1.3\%$ at $A=3$, and $3\%$ at the
extreme $A=5$. This residual is a numerical (grid) effect rather than a physical one: as
$\beta_y^{*}$ is reduced the vertical waist $\sigma_y^{*}$ shrinks below the vertical cell
size, so the cloud-in-cell deposition smooths the peak charge density at the waist---precisely
where the overlap integral $\sum_{ij}Q^{(1)}_{ij}Q^{(2)}_{ij}$ is largest---and the luminosity
is under-counted. The effect is deterministic, not statistical: the run-to-run scatter stays
$\lesssim0.5\%$, far below the bias, and the bias is controlled entirely by the vertical
resolution. Doubling it from $N_y=50$ to $100$ cut the $A=5$ deficit from $8.8\%$ (three
$50\times50\times50$ runs, \ref{app:benchmap}) to $3.0\%$, confirming that the high-$A$ shortfall is set by the vertical sampling of the
waist and not by the physics; sub-percent accuracy at the most strongly focused points is
therefore simply a matter of further vertical refinement. In any case the values $A\gtrsim2$
lie well beyond those of any realistic collider.

%----------------------------------------------------------------
\subsection{Crossing angle}\label{sec:crossingbench}

\begin{figure}[htbp]
\centering
\includegraphics[width=0.6\linewidth]{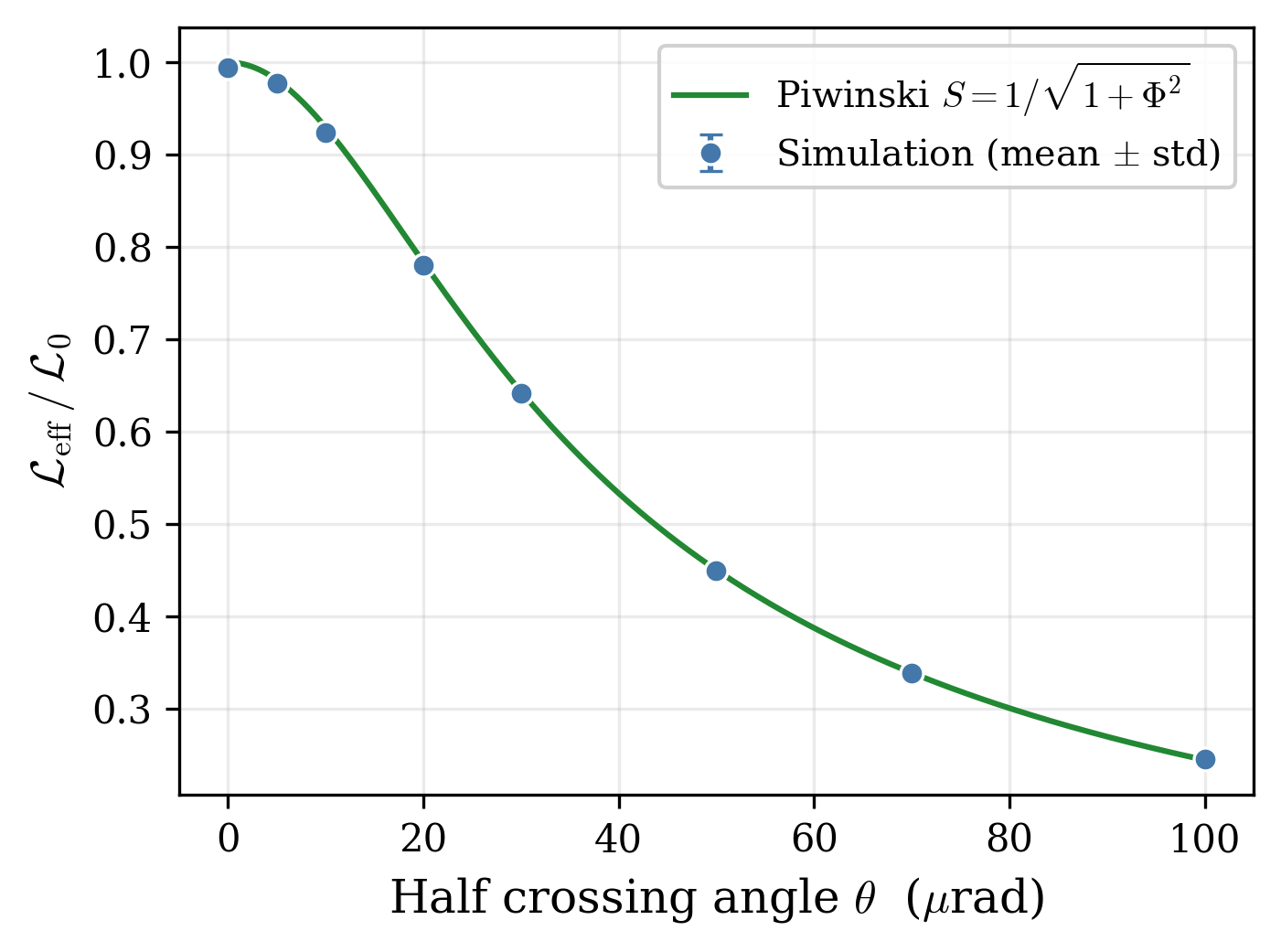}
\caption{Crossing-angle luminosity reduction for symmetric ILC $e^{+}e^{-}$: simulation
(points) and the analytic Piwinski factor $S=1/\sqrt{1+\Phi^{2}}$ (curve) versus the half
crossing angle $\theta$.}
\label{fig:crossing}
\end{figure}

The finite-crossing-angle implementation of Section~\ref{sec:numeric}, the rigid rotation of
Eq.~\eqref{eq:crossing}, is validated against the
analytic Piwinski reduction $S=1/\sqrt{1+\Phi^{2}}$, $\Phi=(\sigma_z/\sigma_y)\tan\theta$ of
Section~\ref{sec:analysis}. Sweeping the half crossing angle $\theta$ for symmetric ILC
$e^{+}e^{-}$ (pinch off, $50^3$ grid, $5\times10^{4}$ macroparticles, mean of three runs), the
simulated $\mathcal{L}_{\rm eff}/\mathcal{L}_{0}$ tracks the Piwinski factor across the full
reduction from $S\simeq1$ down to $S\simeq0.25$ (Fig.~\ref{fig:crossing}), to within
$\sim\!1\%$. The agreement holds up to $\theta\sim10^{-4}$~rad for these beam parameters;
larger angles rotate the longitudinal extent into the vertical plane beyond the transverse
simulation box ($\pm k\sigma_y$) and lie outside the present validation.

%----------------------------------------------------------------
\subsection{Beam--beam pinch: sign verification and interpolation scheme}\label{sec:pinch}

The beam--beam module is verified through the physical sign of the pinch acceleration of
Eq.~\eqref{eq:pinchacc}. Same-charge beams
should defocus (reducing the luminosity), whereas opposite-charge beams should focus
(enhancing it), and the magnitude of the effect should grow with the disruption
parameter~\cite{hollebeek}
$D_y = 2Z N r_0\sigma_z/[\gamma\,\sigma_y(\sigma_x+\sigma_y)]$ (with $Z$ the charge number of the opposing bunch's particle), i.e.\ as the beams become
less energetic and denser (for unit charge $r_0/\gamma=e^{2}/(4\pi\varepsilon_0 E)$, so the
particle mass cancels and only the beam energy $E$ and the bunch geometry remain). Table~\ref{tab:sign} confirms all three
expectations from a single, unmodified code path. The high-energy, large-spot proton beams of
the LHC~\cite{lhc} ($pp$, same-charge) have a vanishing disruption ($D_y\approx0.007$) and are essentially
unmodified, $\mathcal{L}_{\rm eff}/\mathcal{L}_{\rm nom}=0.984$---within $1\%$ of the
pinch-off LHC value $0.992$ at the same $30^3$ grid (three runs, \ref{app:benchmap}),
confirming that the disruption is negligible. For the much less energetic, tightly focused
$e^{+}e^{-}$ beams of ILC-125 the disruption is large ($D_y\approx25$) and the effect is
correspondingly strong and of opposite sign for the two charge combinations: the same-charge
$e^{-}e^{-}$ configuration is defocused to $0.43$, while the opposite-charge
$e^{+}e^{-}$ configuration is focused to $2.22$. The two ILC runs are identical in every
parameter except the sign of the charge product, so this contrast isolates the pinch sign
unambiguously.

Table~\ref{tab:sign} also includes the opposite-charge HERA and EIC configurations of
Section~\ref{sec:asym}. Together with the three ILC and LHC runs above this gives five
collisions spanning more than three orders of magnitude in disruption---same-charge $pp$
($D_y\approx0.007$) and $e^{-}e^{-}$ ($D_y=24.8$), and opposite-charge HERA ($D_y=0.147$),
EIC ($D_y=1.34$) and $e^{+}e^{-}$ ($D_y=24.8$)---against which the magnitude of the pinch
can be checked. For each run the pinch ratio of Table~\ref{tab:sign} is the factor by which
switching the pinch on changes the luminosity with everything else held fixed, so the ratio
minus one is the luminosity change due to the pinch alone: $+1.4\%$ for HERA ($1.014$),
$+6.8\%$ for the EIC ($1.068$) and $+122\%$ for $e^{+}e^{-}$ ($2.22$), an enhancement that
grows monotonically with $D_y$.

\begin{table}[htbp]
\centering
\caption{Sign verification of the beam--beam pinch, and consistency of its magnitude with
$D_y$. The $e^{-}e^{-}$ and $e^{+}e^{-}$ runs (ILC-125, $50^3$) differ only in the charge of
one beam; all runs use $5\times10^4$ macroparticles. For $pp$, $e^{-}e^{-}$ and $e^{+}e^{-}$ (hourglass never enabled) the ratio is
$\mathcal{L}_{\rm eff}/\mathcal{L}_{\rm nom}$ against the flat, hourglass-off analytic
baseline; for HERA and EIC (always run with hourglass enabled, Table~\ref{tab:asym}) the
ratio instead isolates the pinch alone,
$\mathcal{L}_{\rm eff}(\text{hourglass+pinch})/\mathcal{L}_{\rm eff}(\text{hourglass only})$.
Both are ``turn the pinch on, holding everything else fixed'' measurements, so despite the
differing denominators they remain directly comparable as an order-of-magnitude trend in
$D_y$, though not on an identical numerical footing.}
\label{tab:sign}
\resizebox{\textwidth}{!}{%
\begin{tabular}{llccc}
\toprule
Collision & Charge product & $D_y$ & Pinch ratio & Effect \\
\midrule
$pp$ (LHC, $30^3$)         & same $(+,+)$   & $0.0066$ & $0.984$ & negligible (high energy, large spot) \\
HERA ($ep$, $50^3$)        & opposite $(+,-)$ & $0.147$  & $1.014$ & focusing (enhancement) \\
EIC ($ep$, $50^3$)         & opposite $(+,-)$ & $1.34$   & $1.068$ & focusing (enhancement) \\
$e^{-}e^{-}$ (ILC, $50^3$) & same $(-,-)$   & $24.8$   & $0.43$  & defocusing (reduction) \\
$e^{+}e^{-}$ (ILC, $50^3$) & opposite $(-,+)$ & $24.8$   & $2.22$  & focusing (enhancement) \\
\bottomrule
\end{tabular}%
}
\end{table}

Using the physical $e^{+}e^{-}$ configuration we then compare the bicubic and bilinear force
interpolation at a $50^3$ grid with the pinch enabled (Table~\ref{tab:interp}). The two
schemes yield luminosities that differ by $0.7\%$, comparable to the run-to-run scatter, so the bicubic
reconstruction introduces no significant bias. Its cost is purely computational: the
pinch-force interpolation is about $2.2$ times more expensive ($603$~ms versus $277$~ms). This overhead is the price for the
$C^{1}$ force field discussed in Section~\ref{sec:numeric}.

\begin{table}[htbp]
\centering
\caption{Bicubic versus bilinear force interpolation ($e^{+}e^{-}$, ILC-125, $50^3$ grid,
$5\times10^4$ macroparticles, pinch enabled). Values are means of three runs; the bilinear
row was obtained before a correction of the bilinear interpolation weights (cross-check in
\ref{app:benchmap}).}
\label{tab:interp}
\begin{tabular}{lccc}
\toprule
Interpolation & $\mathcal{L}_{\rm eff}$ [$\mathrm{cm^{-2}s^{-1}}$] &
$\mathcal{L}_{\rm eff}/\mathcal{L}_{\rm nom}$ & Pinch-phase time \\
\midrule
Bicubic (Catmull--Rom) & $8.394\times10^{33}$ & $2.217$ & $603$~ms \\
Bilinear               & $8.334\times10^{33}$ & $2.201$ & $277$~ms \\
\bottomrule
\end{tabular}
\end{table}

The enhancement factor $H_L = \mathcal{L}_{\rm eff}/\mathcal{L}_{\rm nom} = 2.22$ and
the disruption parameter $D_y = 24.8$ obtained here for the ILC-125 configuration can be
compared with the ILC design values, which were obtained with GUINEA-PIG and
CAIN~\cite{guineapig,cain}. For the same parameter set ($N=2\times10^{10}$,
$\beta^{*}_{x,y}=13/0.41$~mm, $\gamma\varepsilon_{x,y}=10~\mu\mathrm{m}/35$~nm,
$\sigma_z=0.3$~mm) the ILC parameter table of the 2017 staging report~\cite{ilcstaging} lists
$\mathcal{L}_{\rm geo}=0.374\times10^{34}$ and
$\mathcal{L}=0.82\times10^{34}~\mathrm{cm^{-2}s^{-1}}$, i.e.\ $H_D=2.19$, with $D_y=24.5$
and $\sigma_y^{*}=7.66$~nm; AloHEP gives $\mathcal{L}_{\rm nom}=0.379\times10^{34}~\mathrm{cm^{-2}s^{-1}}$,
$H_L=2.22$, $D_y=24.8$ and $\sigma_y^{*}=7.57$~nm. The geometric quantities differ only
through $\beta_y^{*}=0.40$~mm in the AloHEP parameter set. The two enhancement factors are
not strictly comparable, however: the design $H_D$ includes the hourglass effect, which alone
reduces the luminosity by $\sim\!10\%$ at the ILC value $A\simeq0.75$
(Fig.~\ref{fig:hourglass}), and beamstrahlung, both of which are disabled in this run. With
the hourglass enabled as well (five settings, three runs each: $50^{3}$ with $5\times10^{4}$
and with $5\times10^{5}$ macroparticles, $50\times100\times50$, $50\times50\times100$ and,
with $k=6$, $75\times75\times50$; \ref{app:benchmap}) AloHEP gives $H_L=1.46$--$1.48$
($1.47$ at the standard $50^{3}$, $5\times10^{4}$ setting): the combined enhancement is
smaller than the product of the separate pinch and hourglass factors
($2.22\times0.90\simeq2.0$) and well below the design value of $2.19$. Doubling the vertical
resolution changes the result by $0.5\%$; doubling the number of longitudinal slices, which
halves the time step $\Delta t$ of the velocity-first (semi-implicit) Euler push of
Fig.~\ref{fig:flowchart}, changes it by $0.1\%$; enlarging the $\pm k\sigma$ simulation box
and the truncation of the initial distribution from $k=4$ to $k=6$ at the same cell size
(Section~\ref{sec:numeric}; in this strongly disrupted regime, $D_y\approx25$, the code
reports macroparticles leaving the $k=4$ box, where they receive no force and are not
deposited on the charge grid) lowers it by $0.7\%$; and a tenfold increase of the number of
macroparticles changes it by $0.3\%$. The result is therefore converged to better than $1\%$
with respect to every parameter of the discretisation, so the gap to the design value is a
property of the slice-based model itself, which has not been validated at $D_y\gg1$, a regime
far beyond the $D_y\lesssim1$ of the lepton--hadron colliders that are the target of the code. The pinch-only result shows that the
sign and the order of magnitude of the $e^{+}e^{-}$ pinch are reproduced, but it is obtained
with a rigid (zero-divergence) initial beam, and its closeness to the design value should not
be read as a validation of the strongly disrupted regime. The distinguishing feature of the
code is the extension to the asymmetric regime examined in Section~\ref{sec:asym}, where
$D_y\le1.3$.

%----------------------------------------------------------------
\subsection{Asymmetric application: HERA and EIC}\label{sec:asym}

Finally we apply AloHEP to its target problem: intrinsically asymmetric electron--proton
colliders, in which the two beams differ in species, energy and transverse size. We simulate
HERA ($p\,920\;\mathrm{GeV} \times e\,27.6\;\mathrm{GeV}$, $\sqrt{s}=318.7$~GeV, beam
parameters of the HERA-II configuration as tabulated by the Particle Data
Group~\cite{pdg}) and a
representative EIC configuration ($p\,275\;\mathrm{GeV} \times e\,10\;\mathrm{GeV}$,
$\sqrt{s}=105$~GeV, beam parameters of the EIC Conceptual Design Report~\cite{eiccdr} as
tabulated by the Particle Data Group~\cite{pdg}) on a $50^3$ grid with $5\times10^4$
macroparticles for both machines.
Each collider is run twice (Table~\ref{tab:asym}): with only the hourglass effect enabled
(pinch off), and with the hourglass effect and the beam--beam pinch together, the realistic
configuration.
This separation matters because the available analytic reference---the
unequal-beam generalisation of Eq.~\eqref{eq:Hflat},
\begin{equation}\label{eq:Hunequal}
H=\frac{1}{\sqrt{2\pi}}\int_{-\infty}^{\infty}e^{-u^{2}/2}\,
\frac{\sqrt{\sigma_{x1}^{*2}+\sigma_{x2}^{*2}}\;\sqrt{\sigma_{y1}^{*2}+\sigma_{y2}^{*2}}}
{\sqrt{\sum_{i=1,2}\sigma_{xi}^{*2}\bigl(1+s^{2}/\beta_{xi}^{*2}\bigr)}\;
 \sqrt{\sum_{i=1,2}\sigma_{yi}^{*2}\bigl(1+s^{2}/\beta_{yi}^{*2}\bigr)}}\,du,
\end{equation}
with $s=u\sqrt{\sigma_{z1}^{2}+\sigma_{z2}^{2}}/2$, in which each beam's transverse size
follows its own $\beta$-function and the collision points are spread with r.m.s.\
$\sqrt{\sigma_{z1}^{2}+\sigma_{z2}^{2}}/2$ (for identical beams it reduces to
Eq.~\eqref{eq:Hflat})---captures only the hourglass
reduction; comparing it against a simulation that also includes the pinch would mix two
different physical setups. The hourglass-only runs isolate exactly the physics of the
analytic curve, giving a genuine matched-environment verification, while the hourglass+pinch
runs show the additional, physically real contribution of the beam--beam interaction on top.
The EIC design collides the beams at a $25$~mrad crossing angle compensated by crab
cavities; we simulate the crab-compensated, effectively head-on collision, since the
rigid-rotation model of Section~\ref{sec:numeric} does not include crab kicks and a
$25$~mrad rotation would carry the long proton bunch outside the transverse force grid.
The beam sizes and linear beam--beam tune shifts $\xi_{x,y}$ are listed in
Table~\ref{tab:asympar}.

For beams of unequal transverse size the geometric luminosity must be evaluated with the
\emph{convolved} beam sizes,
\begin{equation}
\mathcal{L}_{\rm geo}=\frac{N_1 N_2 f_c}{2\pi\sqrt{\sigma_{x1}^2+\sigma_{x2}^2}\,
\sqrt{\sigma_{y1}^2+\sigma_{y2}^2}},
\label{eq:lgeo}
\end{equation}
which reduces to Eq.~\eqref{eq:lumgeo0} for identical beams.\footnote{The code's internal nominal-luminosity
readout uses the equal-beam approximation $\mathcal{L}_{\rm nom}=N_1 N_2 f_c/
[4\pi\max(\sigma_{x1},\sigma_{x2})\max(\sigma_{y1},\sigma_{y2})]$, which coincides with
Eq.~\eqref{eq:lgeo} for symmetric beams but underestimates the geometric luminosity when the
beams differ; we therefore compare against Eq.~\eqref{eq:lgeo} for the asymmetric cases.}
Table~\ref{tab:asym} compares the simulated luminosity $\mathcal{L}_{\rm eff}$ with this
analytic baseline.

With the pinch disabled, the simulated ratio reaches $\mathcal{L}_{\rm eff}/\mathcal{L}_{\rm geo}
=0.977$ for HERA and $0.912$ for EIC, within $0.3\%$ and $0.5\%$ respectively of the
analytic hourglass-only prediction of Eq.~\eqref{eq:Hunequal} ($0.980$ and $0.916$): a direct,
matched-environment verification that the PIC engine reproduces the two-beam hourglass
reduction for machines of very different species, energy and beam size. The residual
few-percent deficits are geometric in origin, from the long proton bunches
($\sigma_z\simeq8.5$~cm for HERA, $\simeq6$~cm for EIC) producing a genuine hourglass
reduction that is more pronounced at the EIC, where $\sigma_z/\beta_y^{*}\simeq0.83$ is
comparable to a linear collider.

\begin{figure}[htbp]
\centering
\includegraphics[width=0.6\linewidth]{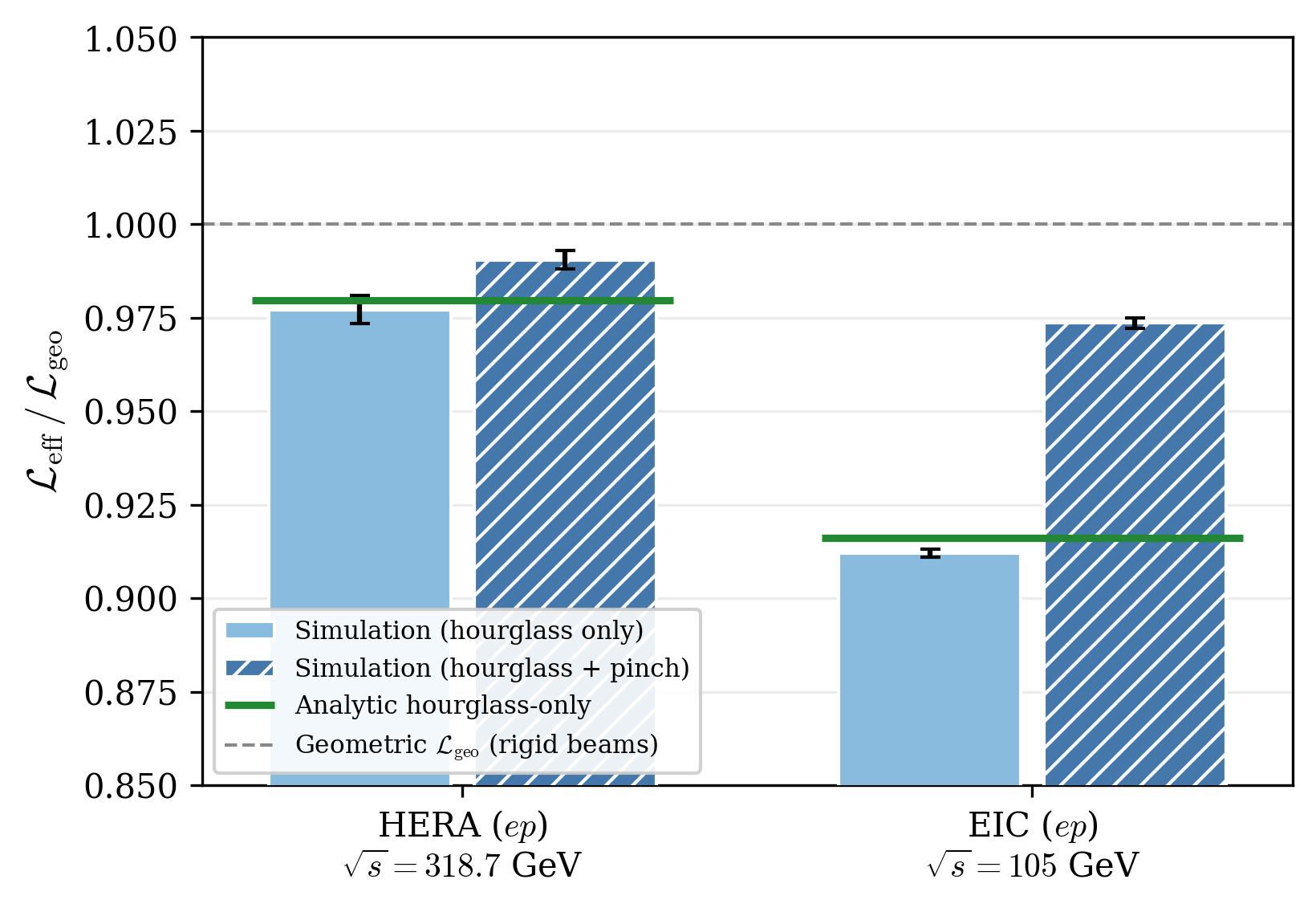}
\caption{Asymmetric electron--proton colliders: simulated $\mathcal{L}_{\rm eff}/\mathcal{L}_{\rm geo}$
for HERA and EIC (bars, mean $\pm$ standard deviation of three runs), each shown as a
matched pair --- hourglass only (light, plain) and hourglass with pinch (dark, hatched) ---
against a single analytic reference, the unequal-beam hourglass factor of Eq.~\eqref{eq:Hunequal} (green line), in
which each beam's transverse size follows its own $\beta$-function. The hourglass-only bar
uses exactly the physics of the analytic curve, so the two are directly comparable; the
hourglass+pinch bar shows the additional, physically real contribution of the beam--beam
pinch, which the analytic curve does not include. The dashed line marks the rigid-beam
geometric luminosity $\mathcal{L}_{\rm geo}$ of Eq.~\eqref{eq:lgeo}.}
\label{fig:asym}
\end{figure}

Enabling the pinch on top of the hourglass raises the ratio to $0.991$ for HERA and $0.974$
for EIC---enhancements of $1.4\%$ and $6.8\%$, respectively, over the hourglass-only
baseline; Table~\ref{tab:sign} in Section~\ref{sec:pinch} shows these two values sit smoothly
on the same $D_y$-dependent trend as the paper's other pinch benchmarks, rather than being an
anomaly specific to the EIC. The difference in magnitude follows directly from the beam parameters through the
disruption parameter and tune shifts of the electron beam, both evaluated with the opposing
proton bunch's full transverse and longitudinal $\boldsymbol{\sigma}$ vector via the code's
\texttt{getDisruption} routine. At HERA the proton spot is large ($111\times30~\mu$m,
Table~\ref{tab:asympar}) and the electron energy high ($27.6$~GeV), so the tune shifts are
small ($\xi_{x,y}\simeq0.022,\,0.036$; $D_y\simeq0.15$), consistent with the near-negligible
pinch contribution there. At the EIC both beams are flat ($\sigma_x/\sigma_y\simeq11$), the
vertical proton spot is $3.5$ times smaller ($8.6~\mu$m) and the electron Lorentz factor
$2.8$ times lower, which gives correspondingly larger tune shifts
($\xi_{x,y}\simeq0.072,\,0.100$; $D_y\simeq1.34$) and a beam--beam pinch that is no longer
negligible, enhancing the luminosity by an amount comparable to the hourglass deficit it sits
on top of. This shows that for electron--proton colliders the balance between
geometric/hourglass and beam--beam corrections is set by the disruption parameter of the
lepton beam, i.e.\ by the hadron spot size and the lepton energy: HERA's large spots leave the
pinch a minor correction, while the EIC's much smaller vertical spot brings it back into play
alongside the hourglass, in the same direction as (though far weaker than) the equal-mass
lepton pinch of Section~\ref{sec:pinch}.
Figure~\ref{fig:asym} summarises this comparison: for both machines the hourglass-only
simulation matches the analytic unequal-beam hourglass factor to within about $1\%$, a
matched-environment verification of the PIC engine's treatment of the hourglass effect in
isolation, while the hourglass+pinch bars show the additional, machine-dependent contribution
of the beam--beam interaction on top. The rigid-beam geometric value $\mathcal{L}_{\rm geo}$
(dashed line) overestimates the luminosity in all cases by neglecting the hourglass effect
entirely.

\begin{table}[htbp]
\centering
\caption{The linear beam--beam tune shifts $\xi_{x,y}$ are those of the (disrupted) electron
beam, from the standard per-plane formula
$\xi_t = |Z_{p}|\,N_{p}\,r_{e}\,\beta^{*}_{e,t}/[2\pi\,\gamma_{e}\,\sigma_{p,t}(\sigma_{p,x}+\sigma_{p,y})]$
($t=x,y$); these are the values emitted by the headless \texttt{-fom} output. Being
analytic functions of the beam parameters, $D_y$ and $\xi_{x,y}$ are deterministic and
independent of the macroparticle sampling (unlike $\mathcal{L}_{\rm eff}$, which is a
Monte-Carlo average over three runs). The GUI \texttt{getBeamBeam} readout uses the same
per-plane formula.}
\label{tab:asympar}
\resizebox{\textwidth}{!}{%
\begin{tabular}{lccccc}
\toprule
Collider & $\sqrt{s}$ [GeV] & proton $\sigma_x\times\sigma_y$ &
electron $\sigma_x\times\sigma_y$ & $\xi_x,\ \xi_y$ (e) & $D_y$ (e) \\
\midrule
HERA ($ep$) & $318.7$ & $111\times30~\mu$m & $110\times30~\mu$m & $0.0224,\ 0.0358$ & $0.1471$  \\
EIC ($ep$)  & $105$ & $94.9\times8.58~\mu$m & $94.8\times8.63~\mu$m & $0.07245,\ 0.09967$ & $1.3420$ \\
\bottomrule
\end{tabular}%
}
\end{table}

\begin{table}[htbp]
\centering
\caption{Asymmetric electron--proton colliders ($50^3$ grid, $5\times10^4$ macroparticles
each; means of three runs). Each collider is run twice: with only the hourglass effect
enabled (pinch off), which is directly comparable to the analytic unequal-beam hourglass
factor of Eq.~\eqref{eq:Hunequal} since both use the same physics, and with hourglass and
pinch together, the realistic configuration. $\mathcal{L}_{\rm geo}$ is the analytic
unequal-beam value of Eq.~\eqref{eq:lgeo}; luminosities in $\mathrm{cm^{-2}s^{-1}}$.}
\label{tab:asym}
\resizebox{\textwidth}{!}{%
\begin{tabular}{llccccc}
\toprule
Collider & $\sqrt{s}$ [GeV] & Configuration & $\mathcal{L}_{\rm geo}$ & $\mathcal{L}_{\rm eff}$ &
$\mathcal{L}_{\rm eff}/\mathcal{L}_{\rm geo}$ & Runtime \\
\midrule
HERA ($ep$) & $318.7$ & hourglass only  & $4.32\times10^{31}$ & $4.22\times10^{31}$ & $0.977$ & $80.0$~s \\
            &         & hourglass+pinch & $4.32\times10^{31}$ & $4.28\times10^{31}$ & $0.991$ & $79.9$~s \\
EIC ($ep$)  & $105$   & hourglass only  & $1.05\times10^{34}$ & $9.55\times10^{33}$ & $0.912$ & $91.4$~s \\
            &         & hourglass+pinch & $1.05\times10^{34}$ & $1.02\times10^{34}$ & $0.974$ & $92.2$~s \\
\bottomrule
\end{tabular}%
}
\end{table}

%----------------------------------------------------------------
\subsection{Coverage across species: heavy ions and muon colliders}\label{sec:species}

\begin{figure}[htbp]
\centering
\includegraphics[width=0.6\linewidth]{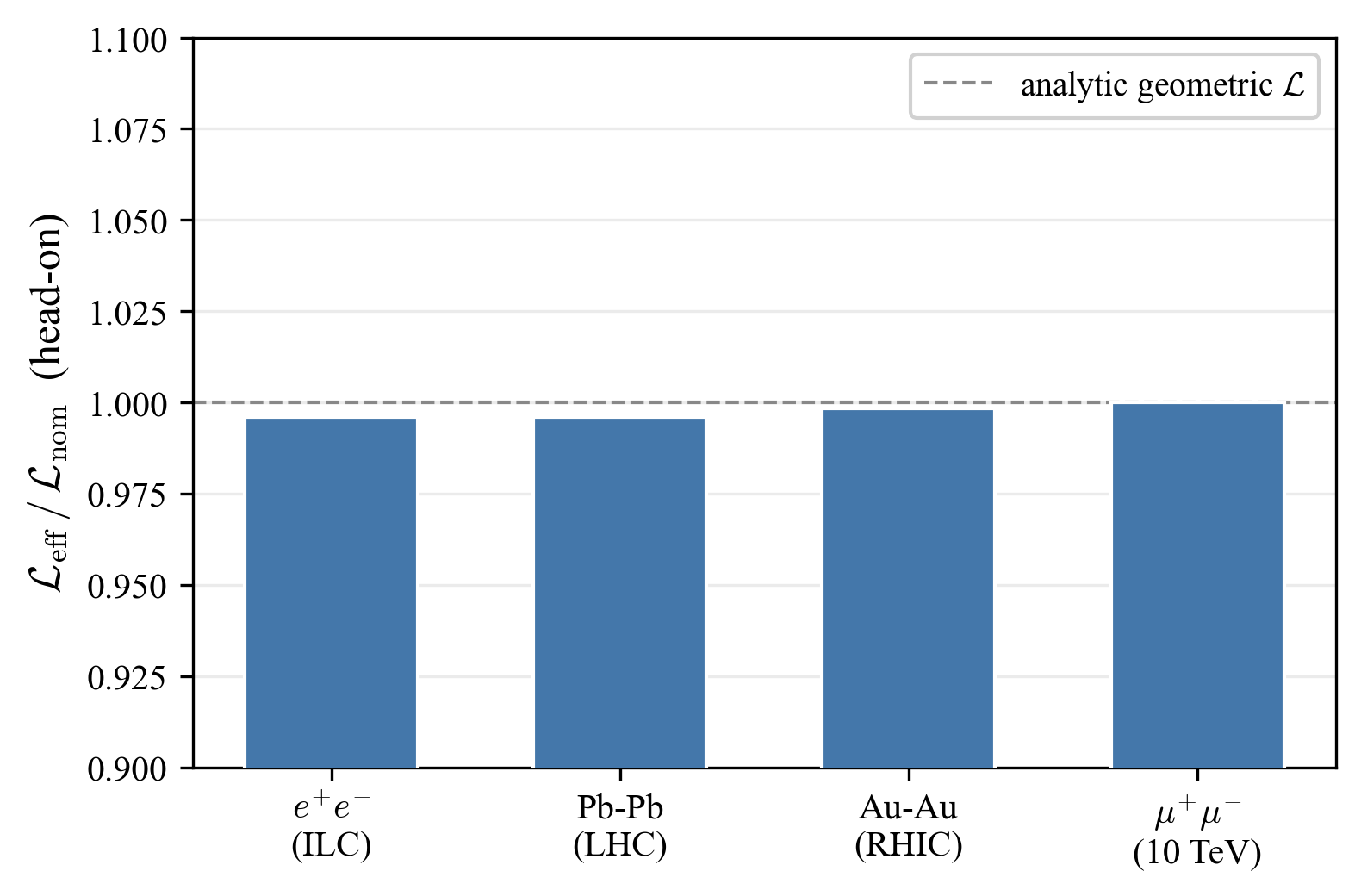}
\caption{Head-on $\mathcal{L}_{\rm eff}/\mathcal{L}_{\rm nom}$ across particle species
($e^{+}e^{-}$, Pb--Pb, Au--Au, $\mu^{+}\mu^{-}$), demonstrating that the PIC engine reproduces
the analytic geometric luminosity independently of species.}
\label{fig:species}
\end{figure}

Because the charge, mass and Lorentz factor of each bunch's species enter only through
the per-bunch constants of Eq.~\eqref{eq:pinchacc} and the kinematics, the same code applies
unchanged to any of the built-in species. We exercise this with two further classes
of machine. For heavy ions we collide Pb--Pb at the LHC ($Z=82$)~\cite{lhc} and Au--Au at RHIC ($Z=79$)~\cite{rhic}
head-on with the pinch disabled; the simulated luminosity reproduces the analytic geometric
value to $\sim\!1\%$, and a Pb--Pb run with the pinch enabled gives
$\mathcal{L}_{\rm eff}/\mathcal{L}_{\rm nom}=0.998$, the expected negligible disruption of
these low-population, high-energy bunches. For a muon collider we collide
$\mu^{+}\mu^{-}$ at $\sqrt{s}=10$~TeV with the consolidated baseline parameters of the
International Muon Collider Collaboration and the MuCol design study~\cite{muon}
($1.8\times10^{12}$ muons per bunch, $\beta^{*}_{x,y}=\sigma_z=1.5$~mm, normalised emittance
$25~\mu$m, an $11.4$~km ring and a $5$~Hz repetition rate), each bunch being stored for one
repetition period ($0.2$~s, i.e.\ $5260$ turns) as in the luminosity estimate of
Ref.~\cite{muon}. With muon decay disabled the geometric luminosity is again recovered to
$\sim\!1\%$ (three runs between $0.997$ and $1.003$), while enabling the decay channel
(\texttt{-decay}) reduces the time-integrated luminosity by the muon-survival factor over the
store---a reduction to $0.252$ for this configuration, against the analytic factor
$F_{\rm decay}=0.254$ of Ref.~\cite{muon}---set by the dilated lifetime $\gamma\tau=0.104$~s
relative to the storage time. The decay factor is applied analytically, so the residual
difference is the statistical scatter of the simulated geometric luminosity.
Figure~\ref{fig:species} summarises the head-on
$\mathcal{L}_{\rm eff}/\mathcal{L}_{\rm nom}$ across $e^{+}e^{-}$, Pb--Pb, Au--Au and
$\mu^{+}\mu^{-}$, all consistent with unity.

%----------------------------------------------------------------
\subsection{Beamstrahlung}\label{sec:beamstrahlung}

The beamstrahlung model in AloHEP is, at present, preliminary: it is not included in the
quantitative benchmarks above, and we defer a complete, validated treatment of beamstrahlung
(and the associated coherent pair production) to the forthcoming \textsc{AloHEPy} rewrite.

\section{Conclusion}\label{sec:conclusion}

Proposals for asymmetric colliders, and especially for lepton--hadron machines of high
physics potential, almost always quote luminosities from the analytic geometric
formula, which assumes head-on collisions of identical, rigid Gaussian bunches. By neglecting
the hourglass effect, the crossing angle and the mutual beam--beam deformation---effects that
are both large and coupled in the interaction region---these estimates can depart
appreciably from realistic operating conditions. Trustworthy luminosity figures
for an asymmetric interaction region therefore require numerical simulation.

We have presented AloHEP, an open-source, cross-platform Java application developed for this
purpose. AloHEP combines a well-established, slice-based Particle-in-Cell
model---cloud-in-cell charge assignment, a direct Green's-function solution of the transverse
potential, finite-difference forces and continuous bicubic force interpolation---with a
dynamic JSON parameter database and an interactive graphical interface. The two colliding
beams may be of entirely different species, energies and sizes, and the hourglass effect,
crossing angle, and beam--beam pinch can each be activated independently. We verified the engine against the analytic limit: with all perturbing effects disabled, the simulated luminosity converges to the geometric value as the grid resolution is
increased, reaching agreement at the $0.1\%$ level on a $100^{3}$ grid, with a systematic
grid-smoothing deficit ($\sim\!10\%$ at $10^{3}$, $\sim\!1\%$ at $30^{3}$) at coarser resolution. Benchmarks confirm that the runtime is dominated by the direct potential solve, that a $30^{3}$ run with $5\times10^{4}$
macroparticles completes in roughly twelve seconds on a single thread, and that
high-resolution $100^{3}$ studies remain tractable on workstation hardware. AloHEP thus
places realistic interaction-region modelling within easy reach of the design and teaching
workflow.

Future development will focus on accelerating the potential solve---the present
$\mathcal{O}(N_x^{2}N_y^{2})$ direct convolution can be replaced by an FFT-based Poisson
solver of $\mathcal{O}(N_xN_y\log N_xN_y)$ cost per slice---and on a more complete,
cross-validated treatment of the beamstrahlung photon spectrum against established codes such
as GUINEA-PIG and CAIN. Finally, AloHEPy (newer version of AloHEP is under development) aims to optimize input parameters to maximize the achievable luminosity of proposed colliders with AI assistance.

%================================================================

%================================================================
\section*{Acknowledgements}
The authors thank U. Kaya and B.B. Oner for their contributions to the original AloHEP code, and
S. Sultansoy for his guidance on the physics of lepton--hadron
colliders.

\section*{Funding}
This work was supported by the Scientific and Technological Research Council of T\"urkiye
(T\"UB\.ITAK) under the 1001 programme [grant No.~125F120].

\section*{CRediT authorship contribution statement}
\textbf{B. Da\u{g}l\i:} Conceptualization, Methodology, Software (core simulation engine and
graphical interface), Writing -- original draft (Sections~\ref{sec:analysis}
and~\ref{sec:numeric}), Writing -- review \& editing.
\textbf{Y. Karaca:} Software (headless interface, benchmark and profiling framework, analysis
scripts), Validation, Investigation, Data curation, Visualization, Writing -- original draft
(Program summary, Sections~\ref{sec:intro}, \ref{sec:software}, \ref{sec:benchmarks}
and~\ref{sec:conclusion}, and the appendices), Writing -- review \& editing.

\section*{Declaration of competing interest}
The authors declare no known competing financial interests or personal relationships that could
have appeared to influence the work reported in this paper.

\section*{Declaration of generative AI and AI-assisted technologies in the writing process}
During the preparation of this work the authors used Claude (Opus 5.5 and Fable 5.1) in order to improve the
language and readability of the manuscript and to cross-check the consistency of the text with
the source code and the benchmark data. After using this tool, the authors reviewed and edited
the content as needed and take full responsibility for the content of the publication. No
generative AI tool was used in the development of the AloHEP code.

\section*{Data availability}
AloHEP, together with all benchmark scripts and the raw result files used in this paper, is openly
available under the MIT license at \url{https://github.com/tobbetuyef/AloHEP}.

%================================================================
%================================================================
%================================================================
\appendix
\setcounter{table}{0}
\renewcommand{\thetable}{A.\arabic{table}}

\section{Benchmark data mapping}
\label{app:benchmap}

\noindent
For reproducibility, Table~\ref{tab:benchmap} maps each result table in
Section~\ref{sec:benchmarks} to the corresponding raw CSV file in the
\texttt{bench\_results/} directory of the source repository.

\begin{table}[htbp]
\centering
\caption{Mapping of paper result tables to raw benchmark CSV files.}
\label{tab:benchmap}
\resizebox{\textwidth}{!}{%
\begin{tabular}{lll}
\toprule
Paper table & CSV file & Configuration \\
\midrule
Table~\ref{tab:conv} (grid convergence) & \texttt{results\_grid\_convergence.csv} & ILC $e^{+}e^{-}$, no pinch, $N_g\!=\!10$--$100$ \\
Table~\ref{tab:npscale} ($N_{mp}$ scaling) & \texttt{results\_np\_scaling.csv} & ILC $e^{+}e^{-}$, $30^{3}$, $10^{4}$--$10^{6}$~mp \\
Table~\ref{tab:sign} (sign verification) & \texttt{results\_lhc\_scaling.csv} ($pp$) & LHC $pp$, same-charge, $30^{3}$ \\
 & \texttt{results\_lhc\_nopinch.csv} ($pp$ baseline) & LHC $pp$, pinch off, $30^{3}$ \\
 & \texttt{results\_ee\_minus\_sign.csv} ($e^{-}e^{-}$) & ILC $e^{-}e^{-}$, same-charge, $50^{3}$ \\
 & \texttt{results\_epinu\_sign.csv} ($e^{+}e^{-}$) & ILC $e^{+}e^{-}$, opposite-charge, $50^{3}$ \\
Table~\ref{tab:sign}--\ref{tab:asympar} $D_y$,$\xi$ & \texttt{results\_fom.csv} & deterministic, all 5 configs \\
Fig.~\ref{fig:hourglass} (hourglass) & \texttt{results\_hourglass.csv} & ILC $e^{+}e^{-}$, hourglass sweep \\
 & \texttt{results\_hourglass\_ny50.csv} & $A=5$ at $50\times50\times50$ ($N_y=50$ check) \\
Section~\ref{sec:pinch} (ILC hourglass+pinch) & \texttt{results\_ilc\_hgpinch.csv} & ILC $e^{+}e^{-}$, hourglass and pinch, $50^{3}$ ($5\times10^{4}$ and $5\times10^{5}$~mp), $50\times100\times50$, $50\times50\times100$ and $75\times75\times50$ ($k=6$) \\
Fig.~\ref{fig:crossing} (crossing angle) & \texttt{results\_crossing.csv} & ILC $e^{+}e^{-}$, $\theta$ sweep \\
Table~\ref{tab:interp} (interp.\ comparison) & \texttt{results\_epinu\_sign.csv} & $e^{+}e^{-}$, $50^{3}$, cubic vs bilinear (\texttt{results\_epinu\_bilinear\_fixed.csv}: bilinear cross-check with the corrected interpolation weights) \\
Tables~\ref{tab:asympar}--\ref{tab:asym} (asymmetric) & \texttt{results\_asymmetric.csv} & HERA + EIC $ep$, $50^{3}$: hourglass-only rows from \texttt{run\_asymmetric\_hgonly.sh}, hourglass+pinch rows from \texttt{run\_asymmetric.sh} \\
Fig.~\ref{fig:asym} (asymmetric summary) & \texttt{results\_asymmetric.csv} & same runs; $\mathcal{L}_{\rm eff}/\mathcal{L}_{\rm geo}$ per run, both configurations \\
Fig.~\ref{fig:species} (heavy ion) & \texttt{results\_heavyion.csv} & Pb--Pb LHC, Au--Au RHIC, $50^{3}$ \\
Fig.~\ref{fig:species} (muon) & \texttt{results\_muon.csv} & $\mu^{+}\mu^{-}$ 10~TeV (MuCol baseline), $\pm$\,decay \\
Fig.~\ref{fig:species} ($e^{+}e^{-}$) & \texttt{results\_grid\_convergence.csv} & $50^{3}$ rows of the grid-convergence run \\
\bottomrule
\end{tabular}}
\end{table}

\noindent
The analytic $\mathcal{L}_{\rm geo}$ in Table~\ref{tab:asym} is the convolved-beam value of
Eq.~\eqref{eq:lgeo}; it is \emph{not} a column of \texttt{results\_asymmetric.csv}, which records
the code's equal-beam $\mathcal{L}_{\rm nom}$ (the readout described in the footnote of
Section~\ref{sec:asym}). The two differ only by a purely geometric factor,
\[
\mathcal{L}_{\rm geo}=\mathcal{L}_{\rm nom}\,
\frac{2\max(\sigma_{x1},\sigma_{x2})\max(\sigma_{y1},\sigma_{y2})}
{\sqrt{\sigma_{x1}^2+\sigma_{x2}^2}\,\sqrt{\sigma_{y1}^2+\sigma_{y2}^2}},
\]
so $\mathcal{L}_{\rm geo}$ is recovered directly from the recorded $\mathcal{L}_{\rm nom}$ and the
beam sizes of Table~\ref{tab:asympar}.

%================================================================
%================================================================
\setcounter{table}{0}
\renewcommand{\thetable}{B.\arabic{table}}

\section{Notation}
\label{app:notation}

\noindent
Table~\ref{tab:notation} lists the symbols used in this paper.

\begin{longtable}{@{}llp{0.50\textwidth}@{}}
\caption{Symbols used in this paper.} \label{tab:notation} \\
\toprule
Symbol & Unit & Description \\
\midrule
\endfirsthead
\toprule
Symbol & Unit & Description \\
\midrule
\endhead
\bottomrule
\endfoot
\multicolumn{3}{@{}l}{\textit{Beam and bunch parameters}} \\
$N_1, N_2$ & --- & Bunch populations (real particles per bunch) \\
$N_p$ & --- & Number of macroparticles per bunch \\
$f_c$ & Hz & Bunch collision rate \\
$\sigma_x, \sigma_y$ & m & Transverse bunch sizes (r.m.s.) \\
$\sigma_s, \sigma_z$ & m & Longitudinal bunch size (r.m.s.) \\
$\rho(x,y,s)$ & --- & Normalised particle density function \\
\multicolumn{3}{@{}l}{\textit{Beam optics (Courant--Snyder)}} \\
$\varepsilon$ & m\,rad & Emittance \\
$\alpha, \beta, \gamma$ & --- & Twiss parameters \\
$\beta^{*}$ & m & Beta function at the interaction point \\
$\sigma^{*}$ & m & Transverse beam size at the IP \\
\multicolumn{3}{@{}l}{\textit{Kinematics}} \\
$\sqrt{s}$ & GeV & Centre-of-mass energy \\
$\gamma$ & --- & Lorentz factor ($E/m$) \\
$\theta$ & rad & Half crossing angle (full angle $= 2\theta$) \\
$\Phi$ & --- & Piwinski parameter ($\sigma_z\tan\theta/\sigma_y$) \\
$S$ & --- & Geometric-overlap reduction factor ($1/\sqrt{1+\Phi^2}$) \\
$r_e, r_p$ & m & Classical radius of the beam particle \\
\multicolumn{3}{@{}l}{\textit{Numerical grid}} \\
$N_g$ & --- & Transverse grid resolution (per axis) \\
$N_x, N_y, N_z$ & --- & Grid resolution per axis \\
$\Delta x, \Delta y, \Delta z$ & m & Cell / slice dimensions \\
$\Delta t$ & s & Time step \\
$k$ & --- & Collision-region size coefficient \\
$Q_{ij}$ & --- & Charge-matrix element (cloud-in-cell deposition) \\
$\phi_{mn}$ & --- & Electrostatic potential at grid corner $(m,n)$ \\
$F_x, F_y$ & --- & Force-matrix elements (finite differences) \\
\multicolumn{3}{@{}l}{\textit{Luminosity}} \\
$\mathcal{L}_0$ & cm$^{-2}$s$^{-1}$ & Geometric luminosity (identical beams) \\
$\mathcal{L}_{\rm nom}$ & cm$^{-2}$s$^{-1}$ & Code's equal-beam nominal-luminosity readout, $N_1N_2f_c/(4\pi\,\sigma_x^{\max}\sigma_y^{\max})$ with $\sigma_t^{\max}=\max(\sigma_{t1},\sigma_{t2})$; equals $\mathcal{L}_0$ for identical beams \\
$\mathcal{L}_{\rm geo}$ & cm$^{-2}$s$^{-1}$ & Convolved-beam geometric luminosity \\
$\mathcal{L}_{\rm eff}$ & cm$^{-2}$s$^{-1}$ & Simulated effective luminosity \\
\multicolumn{3}{@{}l}{\textit{Beam--beam figures of merit}} \\
$D_{x,y}$ & --- & Disruption parameters \\
$\xi_{x,y}$ & --- & Linear beam--beam tune shifts \\
\end{longtable}

\end{document}